\documentclass[iop]{emulateapj}

\usepackage[usenames,dvipsnames]{color}
\usepackage{amssymb, amsmath, apjfonts, natbib}
\usepackage{multirow, placeins, caption3, epstopdf}
\usepackage{ulem}

\def\degree{{^{\circ}}}
\newdimen\digitwidth
\setbox1=\hbox{0}
\digitwidth=\wd1
\catcode`"=\active
\def"{\kern\digitwidth}

\def\kms{{\;\mathrm{km}\;\mathrm{s}^{-1}}}
\def\masyr{{\;\mathrm{mas}\;\mathrm{yr}^{-1}}}
\def\kmskpc{{\;\mathrm{km}\;\mathrm{s}^{-1}\;\mathrm{kpc}^{-1}}}
\def\kpc{{\;\mathrm{kpc}}}
\def\pc{{\;\mathrm{pc}}}
\def\deg{{\;\mathrm{deg}}}
\def\Gyr{{\;\mathrm{Gyr}}}

\slugcomment{}

\begin{document}

\title{Kinematics of the X-shaped Milky Way Bulge:\\
       Expectations from a Self-consistent $N$-body Model}

\author{Yujing Qin\altaffilmark{1, 2},
        Juntai Shen\altaffilmark{1, 8},
        Zhao-Yu Li\altaffilmark{1},
        Shude Mao\altaffilmark{3, 4, 5},
        Martin C. Smith\altaffilmark{1},\\
        R. Michael Rich \altaffilmark{6},
        Andrea Kunder\altaffilmark{7},
        and Chao Liu\altaffilmark{3}}

\altaffiltext{1}{Key Laboratory for Research in Galaxies and Cosmology,
                 Shanghai Astronomical Observatory, Chinese Academy of Sciences, 80 Nandan Road, Shanghai 200030, China}
\altaffiltext{2}{University of Chinese Academy of Sciences, 19A Yuquan Road, Beijing 100049, China}
\altaffiltext{3}{National Astronomical Observatories, Chinese Academy of Sciences,
                 20A Datun Road, Chaoyang District, Beijing 100012, China}

\altaffiltext{4}{Physics Department and Tsinghua Center for Astrophysics, Tsinghua University, Beijing, 100084, China}
\altaffiltext{5}{Jodrell Bank Centre for Astrophysics, The University of Manchester, Alan Turing Building, Manchester M13 9PL, UK}
\altaffiltext{6}{Department of Physics and Astronomy, University of California Los Angeles,
                 430 Portola Plaza, Box 951547, Los Angeles, CA 90095-1547}
\altaffiltext{7}{Leibniz-Institute f\"{u}r Astrophysik Potsdam (AIP), An der Sternwarte 16, D-14482 Potsdam, Germany}
\altaffiltext{8}{Correspondence should be addressed to Juntai Shen at jshen@shao.ac.cn}

\begin{abstract}

We explore the kinematics (both the radial velocity and the proper motion) of the vertical X-shaped feature in the Milky Way with an $N$-body bar/bulge model. From the solar perspective, the distance distribution of particles is double-peaked in fields passing through the X-shape. The separation and amplitude ratio between the two peaks qualitatively match the observed trends towards the Galactic bulge. We confirm clear signatures of cylindrical rotation in the pattern of mean radial velocity across the bar/bulge region. We also find possible imprints of coherent orbital motion inside the bar structure in the radial velocity distribution along $l=0\degree$, where the near and far sides of the bar/bulge show excesses of approaching and receding particles. The coherent orbital motion is also reflected in the slight displacement of the zero-velocity-line in the mean radial velocity, and the displacement of the maximum/minimum in the mean longitudinal proper motion across the bulge region. We find some degree of anisotropy in the stellar velocity within the X-shape, but the underlying orbital family of the X-shape cannot be clearly distinguished. Two potential applications of the X-shape in previous literature are tested, i.e., bulge rotation and Galactic center measurements.
We find that the proper motion difference between the two sides of the X-shape can be used to estimate the mean azimuthal streaming motion of the bulge, but not the pattern speed of the bar. We also demonstrate that the Galactic center can be located with the X-shape, but the accuracy depends on the fitting scheme, the number of fields, and their latitudinal coverage.

\end{abstract} 


\section{INTRODUCTION}     

The Galactic bulge holds important information towards understanding the formation and evolution of the Milky Way (MW). However, our edge-on perspective in the disk plane and the large and variable dust extinction especially in optical bands make it a non-trivial task to study the MW bulge. Despite these difficulties, our knowledge of the Galactic bulge has advanced greatly in the past two decades. Numerous studies indicate that the MW bulge contains a bar structure. For example, the non-circular motions of gas in the inner Galaxy was explained with bar perturbation (de Vaucouleurs 1964; Fux 1999; Englmaier \& Gerhard 1999; Bissantz \& Gerhard 2002). Likewise, the boxy shape of the MW bulge revealed by \textit{COBE/DIRBE} photometry (Dwek et al. 1995) was characterized as the perspective effect of a tilted bar (Blitz \& Spergel 1991; Binney et al. 1997). More evidence for the barred morphology emerges from star counts using distance indicators like Red Clump (RC) stars (Stanek et al. 1994; Rattenbury et al. 2007; Cao et al. 2013). Kinematic information of the MW bulge stellar populations provides additional insights into its origin. The Bulge Radial Velocity Assay (BRAVA) revealed the fast cylindrical rotation of the Galactic bulge extending up to $|b|=8\degree$, which is inconsistent with the classical scenario that the bulge is built in mergers and dominated by random motions (Howard et al. 2008, 2009; Kunder et al. 2012). Shen et al. (2010) demonstrated that a pure-disk galaxy with a spontaneously formed bar structure can match the BRAVA observation of the radial velocity and velocity dispersion towards the bulge region well. They also found that the mass of a classical bulge, if in existence, may be relatively small. Therefore, the Galactic bulge is likely built through bar-induced internal dynamical processes, with minor contribution from a merger-built classical bulge.

Recently, some studies suggested that the Galactic bulge shows the signature of a vertical X-shape. Towards the bulge region, the apparent magnitudes of RCs in certain fields show double-peaked distributions, with increasing peak separations towards higher latitudes (McWilliam \& Zoccali 2010; Nataf et al. 2010; Saito et al. 2011; Ness et al. 2012; Nataf et al. 2015). Since RCs are well-calibrated standard candles with a roughly constant absolute magnitude, such a double-peaked feature seems to indicate the existence of a vertical X-shaped feature in the Galactic bulge. More detailed studies revealed that certain latitudinal slices of the Galactic bulge have two density lumps aligned with the bar, which indeed results in the observed double-peaked apparent magnitude distributions of RCs (Saito et al. 2011). Based on the RCs in the VISTA Variables in the Via Lactea (VVV) Survey, the symmetrized three-dimensional bulge model of Wegg \& Gerhard (2013) also clearly illustrates the X-shape in its side-on view and surface density slices. Further Made-to-Measure modelling reveals an off-centered X-shape in the Galactic bulge, and the peanut/X-shaped component accounts for over $20\%$ of the bulge stellar mass (Portail et al. 2015a). The X-shaped bulge morphology is frequently seen in edge-on disk galaxies (e.g. Whitemore \& Bell 1988; L\"{u}tticke et al. 2000; Bureau et al. 2006), and is presumably similar to the structure in our own Galaxy.

Using a simple and self-consistent $N$-body simulation, Li \& Shen (2012) showed that such an X-shape can arise naturally during the evolution of the Galactic bar. Numerical simulations demonstrated that a bar forms via the bar instability, and then thickens vertically due to the buckling instability (Combes \& Sanders 1981; Raha et al. 1991). The vertical motion in a peanut/X-shape bulge is enhanced after the buckling episode.
Some studies proposed that x1 orbits with 2:1 vertical Lindblad resonance, namely \textit{banana orbits} for their side-on appearance, can contribute to the peanut/X-shaped feature in edge-on galaxies (Pfenniger \& Friedli 1991; Patsis et al. 2002; Athanassoula 2005). Great efforts have been made to understand the roles of banana orbits (and their associated orbits) in forming the peanut/X-shaped bulges (e.g. Patsis \& Katsanikas 2014a, b). However, recent work by Portail et al. (2015b) classified orbital families in the peanut/X-shaped bulges, and suggested the brezel-like orbit as the main contributor to the X-shape. Since all of these orbital families can potentially support the peanut/X-shaped feature, additional kinematic information is required to distinguish their contributions in the Galactic X-shape. If the Galactic X-shape is dominated by certain orbital families, its stellar kinematics is expected to reflect the underlying orbital motion. Thus, the stellar kinematics is essential to better understand the Galactic X-shape.

Currently, only a handful of observational studies have explored the kinematics inside the X-shape. V\'asquez et al. (2013) (hereafter V13) analyzed the stellar motion in the X-shape using RCs in $(l,b)=(0\degree,-6\degree)$. With the three-dimensional velocity available for $352$ stars, they found weak anti-correlation between the longitudinal proper motion and the radial velocity in both bright and faint RCs. Meanwhile, no significant correlation was seen between the latitudinal proper motion and the radial velocity in their sample. They interpreted the correlation with stars on elongated bar orbits, possibly the streaming motion along the arms of the X-shape. Poleski et al. (2013) (hereafter P13) investigated the proper motion of RCs in fields at $b\sim5\degree$. Between the probability-distinguished closer and further arms of the X-shape, the mean proper motion difference is asymmetric in both $l$ and $b$ directions, which is linear for $-0.1\degree<l<0.5\degree$, but roughly constant for $-0.8\degree<l<-0.1\degree$. They attributed the linear side to the streaming motion within the X-shape. From a numerical perspective, Gardner et al. (2014) (hereafter G14) first explored the kinematic properties of X-shaped bulges. They compared the kinematics at the closer and further sides of the bar/bulge, and suggested that X-shaped bugles show coherent signature of a minimum along $l=0\degree$ in the mean radial velocity difference between the closer and further sides. Although efforts have been made in both observations and simulations, the kinematics of the X-shape is still poorly understood.

In this paper, we exploit the power of simulations to present a closer look at the X-shape in our own Galaxy. Using an $N$-body model of the Milky Way-like barred galaxy, we explore the kinematic features of the Galactic X-shape, and compare our model to the observations and other simulation works. The numerical model here can not represent the actual Galactic bulge in all aspects, but its self-consistency and simplicity still reveal the general properties of the X-shaped bulge in such exploration and comparison. Our results lead to a better understanding of the existing observations, serve as an independent check of other numerical works, and make predictions for future observations. In Section 2, we briefly describe the $N$-body model used in this paper. Results and comparisons with existing observations are shown in Section 3. In Section 4, we discuss the nature of the X-shape by analyzing the correlations between proper motions and radial velocities, and test two potential applications to derive Galactic parameters based on the Galactic X-shape. Our results are summarized in Section 5.

\section{MODEL}

In this work, we use the same $N$-body model as in Shen et al. (2010) and Li \& Shen (2012). Initially it contains a cold (Toomre's $Q\sim1.2$) exponential stellar disk of $10^6$ particles rotating in a rigid dark matter halo potential. The bar forms rapidly within the first gigayear, and becomes thickened vertically in the inner region via the buckling instability. After $\sim2.4\Gyr$, the structure of this model reaches a quasi-steady state, with a bar pattern speed of $\sim40\kmskpc$ (Shen 2014; Molloy et al. 2015). The snapshot at $\sim4.8\Gyr$ was analyzed in Shen et al. (2010) and Li \& Shen (2012).

Assuming a distance of the Sun to the Galactic center (GC) ($R_0\sim8.5\kpc$) and a $20\degree$ angle of Sun--GC line to the bar major axis, this model yields excellent matches with the mean radial velocity and velocity dispersion measurements of BRAVA. The X-shape in this model was highlighted after subtracting a smooth component from the edge-on projection (Li \& Shen 2012). The distance distributions in the model towards the bulge fields can qualitatively match the apparent magnitude distributions of RCs in McWilliam \& Zoccali (2010). Thus, despite its simplicity, this model is valuable in understanding the kinematics of the Galactic X-shape.

For consistency, we adopt the same solar configuration as in Shen et al. (2010) and Li \& Shen (2012). Following G14, we also split the particles in a field into two samples to illustrate the kinematic difference between the two sides of the bar/X-shape. Particles within $R_0$ from the Sun are denoted as ``near'', and those beyond $R_0$ are denoted as ``far''. We calculated the heliocentric radial velocity (or line-of-sight velocity $V_{\mathrm{los}}$) and proper motion ($\mu_l^{\star}$, $\mu_b$) of particles inside the bar/bulge region (Galactocentric radii less than $4.5\kpc$). Along the minor axis of the Galactic bulge, some specific fields are carefully studied in order to allow for detailed comparisons with observations. Across the bulge region, particles are binned in the $(l,b)$ plane to map the spatial trend of the mean velocity and velocity dispersion. In observations, $\mu_l$ from multi-epoch measurements needs a factor $\cos b$ to represent its angular speed in the longitudinal direction, as required by the metric of spherical coordinates. In this paper we also use $\mu_l^{\star} = \mu_l \cos b$ as the notation of the metric-corrected longitudinal proper motion.

\section{RESULTS AND DISCUSSION}

\subsection{Distance Distribution towards the X-shaped Bulge}

From the solar perspective, the Galactic X-shape features double-peaked magnitude/distance distributions in fields passing through it (Li \& Shen 2012). In this section, we study the distance distributions towards the X-shape in this model in greater detail, and investigate the properties of the double-peaked feature across the bulge region.

We use kernel density estimator to test if the distance distribution is double-peaked (Silverman 1981; Hall \& York 2001). In each field, the distance distribution is constructed from the sample particles with a Gaussian kernel. The multimodality of the distance distribution is illustrated with the critical kernel sizes for $k$ modes ($h_k$), and its significance levels are evaluated using bootstraping. A relatively large and statistically significant $h_k$ implies the existence of more than $k$ modes. For each field, we search for the critical kernel size for a single-peaked or double-peaked distribution ($h_1$ and $h_2$, respectively), and test their significance levels ($p_1$, $p_2$) via standard bootstrapping. The details of the method is described in Appendix A.

To illustrate the properties of the double-peaked feature, we locate the two peaks in the distance distribution using the mean value of $h_1$ and $h_2$ as an optimal kernel size, and find the separation of the two peaks in the distance distribution and the ratio of peak amplitudes (far to near). It is worth pointing out that the algorithm here is relatively conservative, while the visual identification (with \textit{a priori} knowledge about the double peaks) is more flexible in distinguishing modes.

\begin{figure*}[ht!]
\centering
\includegraphics[height=2.2in]{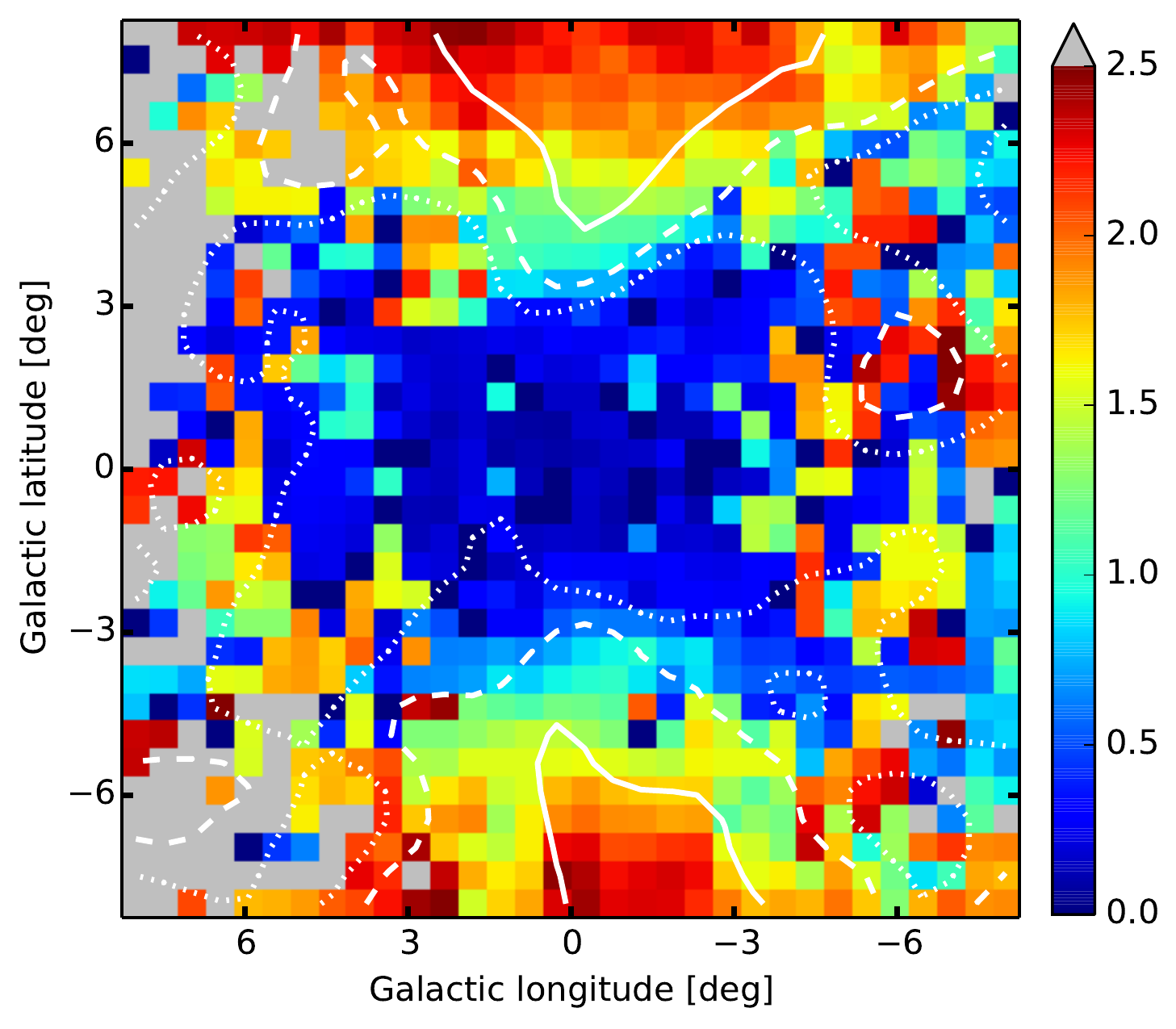}
\includegraphics[height=2.2in]{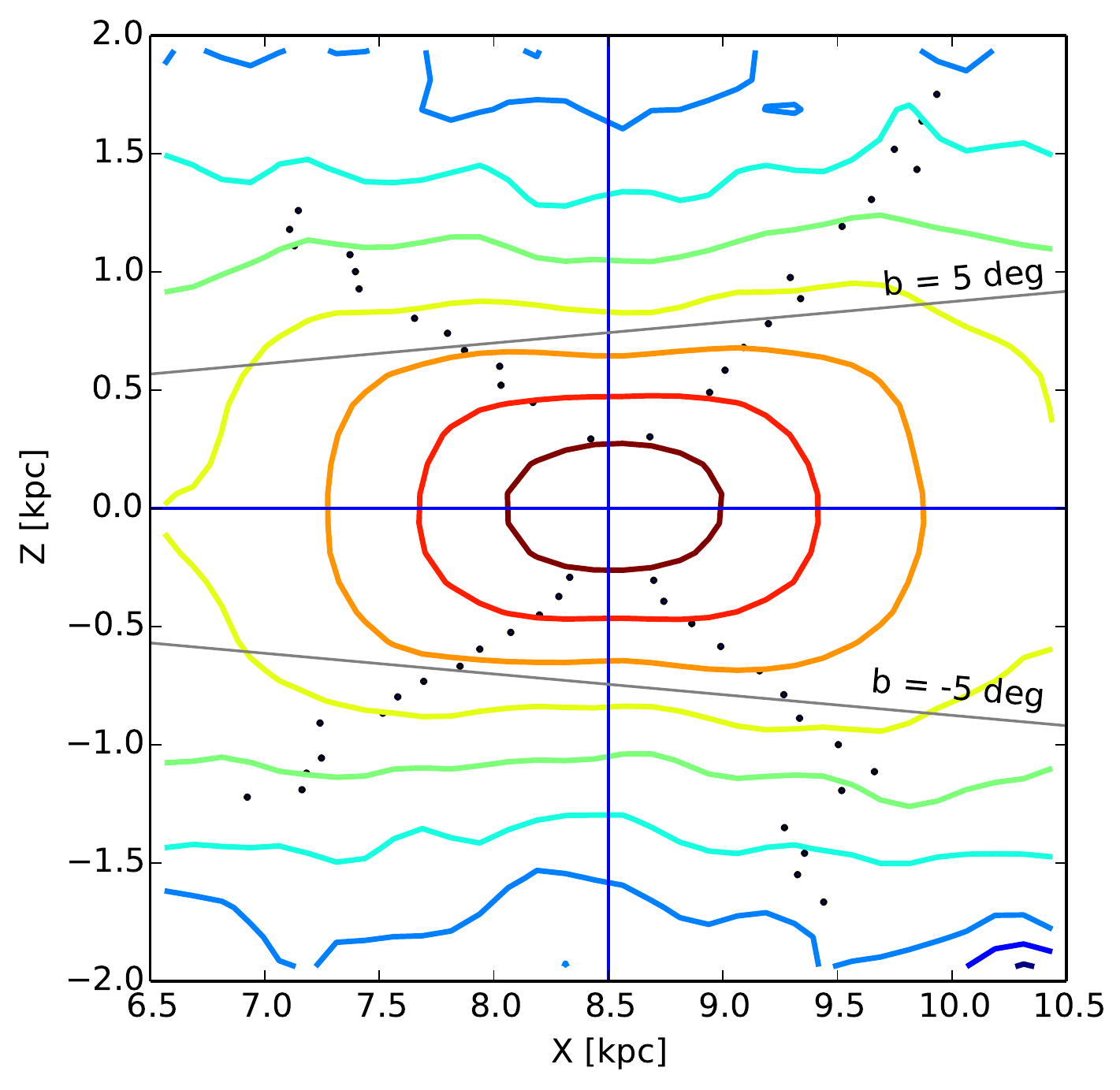}
\\
\includegraphics[height=2.2in]{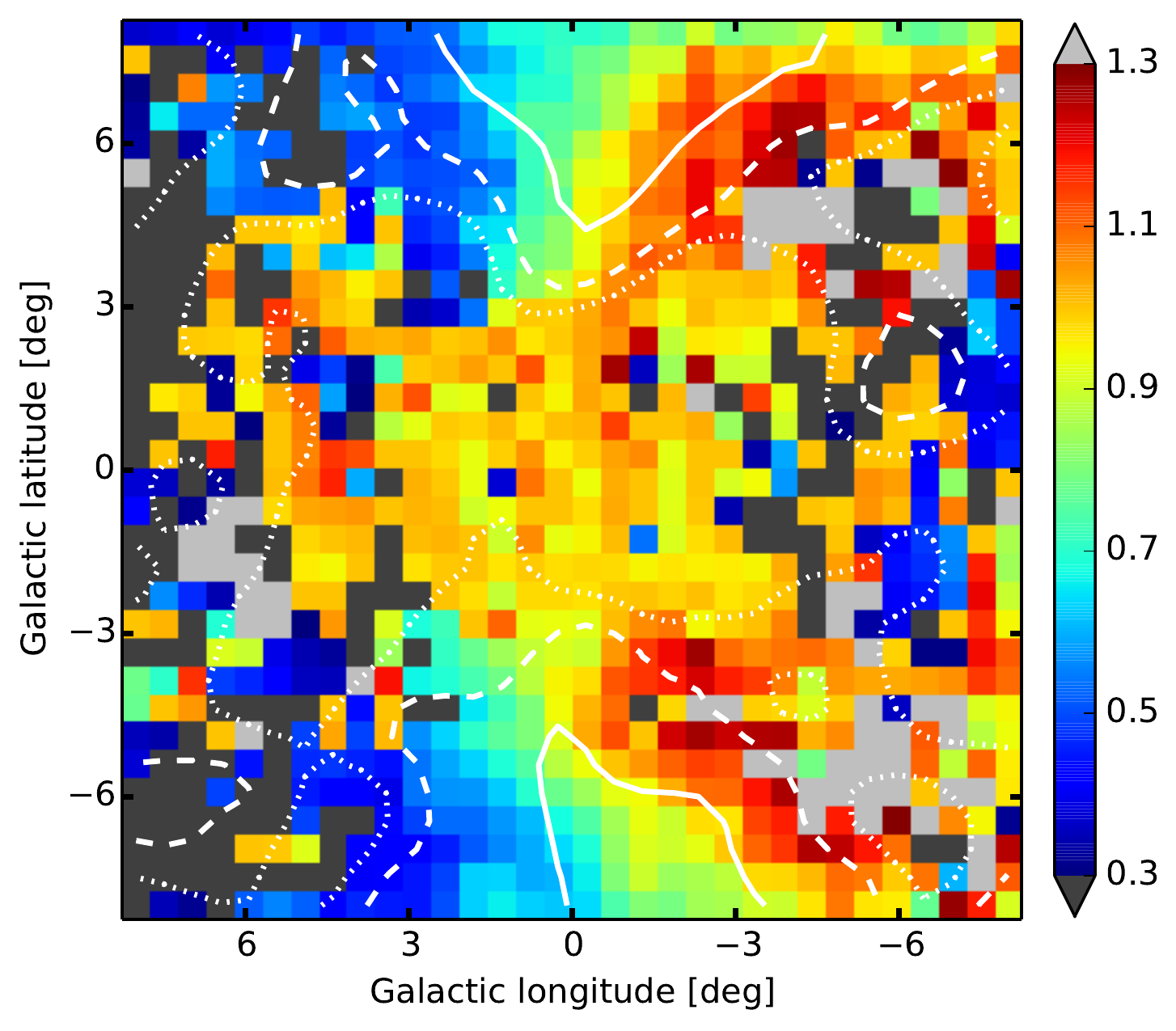}
\includegraphics[height=2.2in]{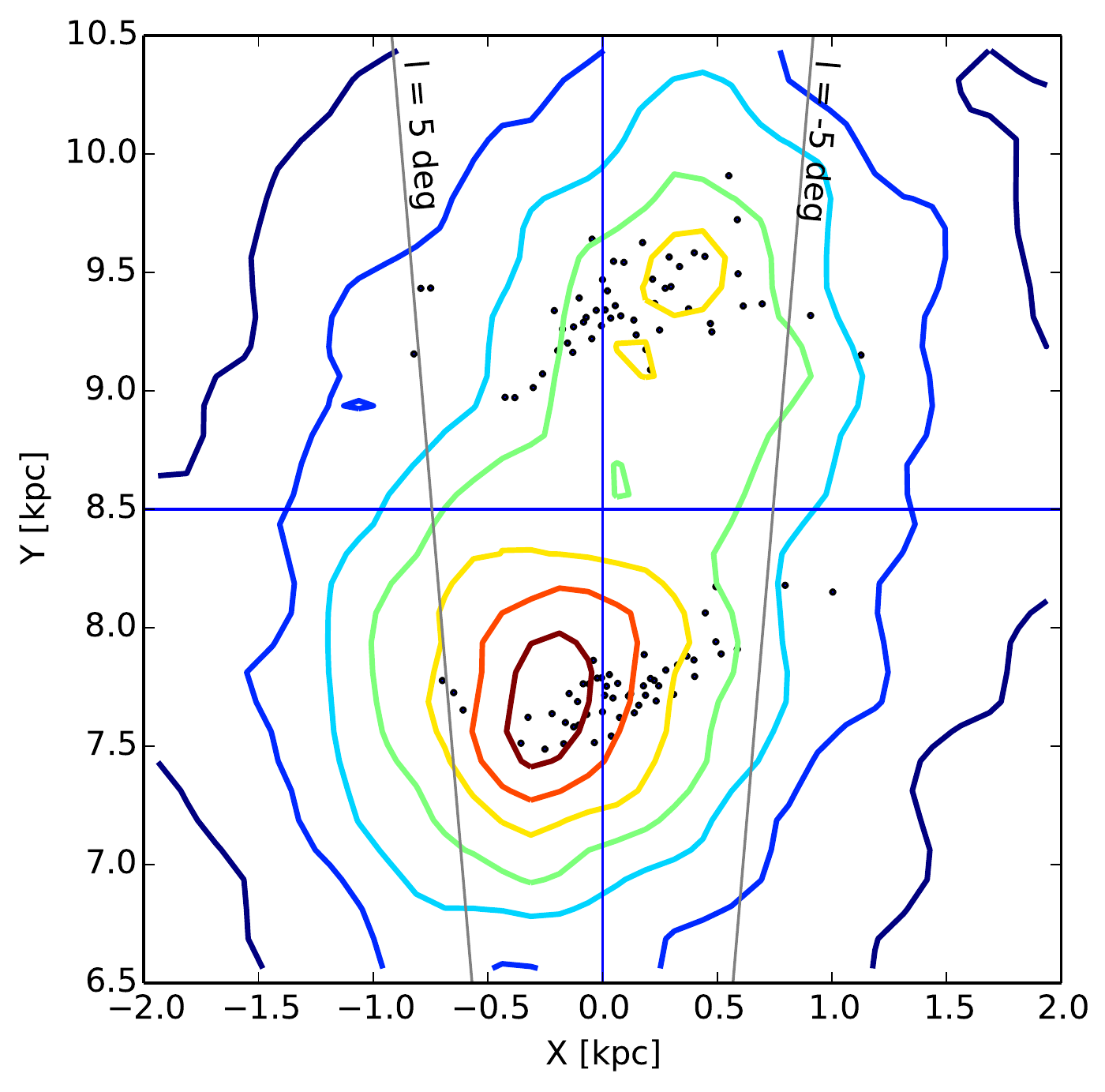}
\caption
{
Geometric properties of the double-peaked feature in the distance distribution. The left column shows the separation (upper panel, in $\kpc$) of the two peaks and their amplitude ratio (lower panel, far to near) across the bulge region. Dotted, dashed and solid contour lines represent $p_1=0.5$, $0.68$ and $0.95$, respectively. Color bars are scaled to highlight the trends in the region where the double-peaked feature is clear. The right column shows the peak positions in the $l=0\degree$ plane (upper panel) and $b=5\degree$ plane (lower panel, XY-projection). Contours display the particle density in logarithmic scale, and the black dots mark the identified peaks with $p_1>0.68$. Lines for $l=\pm 5\degree$ and $b=\pm 5\degree$ are also marked. In the region where the double-peaked feature is clearly identified, the peak separation steadily increases with $|b|$ (upper left), and the peak ratio (far to near) decreases with $l$ (lower left).
} 
\label{fig:double_peak_map}
\end{figure*}

Figure \ref{fig:double_peak_map} shows the properties of the double-peaked distance distributions across the bulge region. Inside the regions where the double-peaked feature is robustly detected, we find some prominent trends. The separation of the two peaks in the distance distribution increases with $|b|$ due to the X-shape geometry (upper left panel), which is especially clear in the vertical slice of $l=0\degree$ (upper right panel). While at the same latitude, the separation is roughly constant. Meanwhile, the ratio of peak amplitude (far to near) decreases with $l$. Since the bar is tilted, its near and far sides dominate at $l>0\degree$ and $l<0\degree$ respectively, as illustrated in the latitudinal slice of $b=5\degree$ (lower right panel). As $l$ decreases, the peak amplitude decreases at the near side but increases at the far side, which further increases the ratio of peak amplitude (far to near). In addition, at both sides, the identified peaks are closer to the observer as $l$ increases (lower right panel). In other regions where the double-peaked feature is not robustly detected (low $p_1$), the separation and peak ratio are spiky since the secondary peaks are likely to be spurs in the under-smoothed distribution there. The spatial distributions of the overdensity shown in the right column of Figure \ref{fig:double_peak_map} are qualitatively consistent with the recent OGLE-III RC observations in Figure 9 of Nataf et al. (2015).

The increasing peak separation with $b$ in this model is qualitatively consistent with previous studies (e.g. McWilliam \& Zoccali 2010; Saito et al. 2011; Li \& Shen 2012; Nataf et al. 2015). The peak separation measured in $(0\degree, 6\degree)$ ($\sim900\;\pc$ above the disk plane at the GC) is about $1.7\kpc$, which is similar to the result in Figure 8 of McWilliam \& Zoccali (2010). The varying peak amplitude due to the tilted bar/X-shape agrees with the results in McWilliam \& Zoccali (2010) and Saito et al. (2011), and naturally leads to the decreasing ratio of peak amplitudes (far to near) with $l$ (lower left panel of Figure \ref{fig:double_peak_map}). Some minor disagreements exist on the latitudinal range where the double-peaked feature is detectable. The two split RCs in McWilliam \& Zoccali (2010) merge around $b\sim5\degree$, and Saito et al. (2011) found a similar but slightly lower limit of $|b|\sim4\degree$. The double-peaked density profile in Wegg \& Gerhard (2013) emerges at $400\;\pc$ above the Galactic plane, corresponding to $b\sim2.7\degree$ at $R_0=8.5\kpc$. In our model, the double-peaked distance distribution can be detected at $|b|\sim3\degree$ ($\sim450\;\pc$ above the disk, at $R_0=8.5\kpc$), which is similar to the result obtained by Wegg \& Gerhard (2013). Saito et al. (2011) obtained a longitudinal range of $|l|\leq2\degree$ for the double-peaked feature, while the range is latitude-dependent in our model. Towards higher latitudes, the longitudinal range for the double-peaked feature in our model becomes much wider than $|l|\leq2\degree$ in Saito et al. (2011). These minor disagreements could be due to the different peak-finding methods adopted.

\begin{figure*}[h!]
\centering
\includegraphics[width=0.38\textwidth]{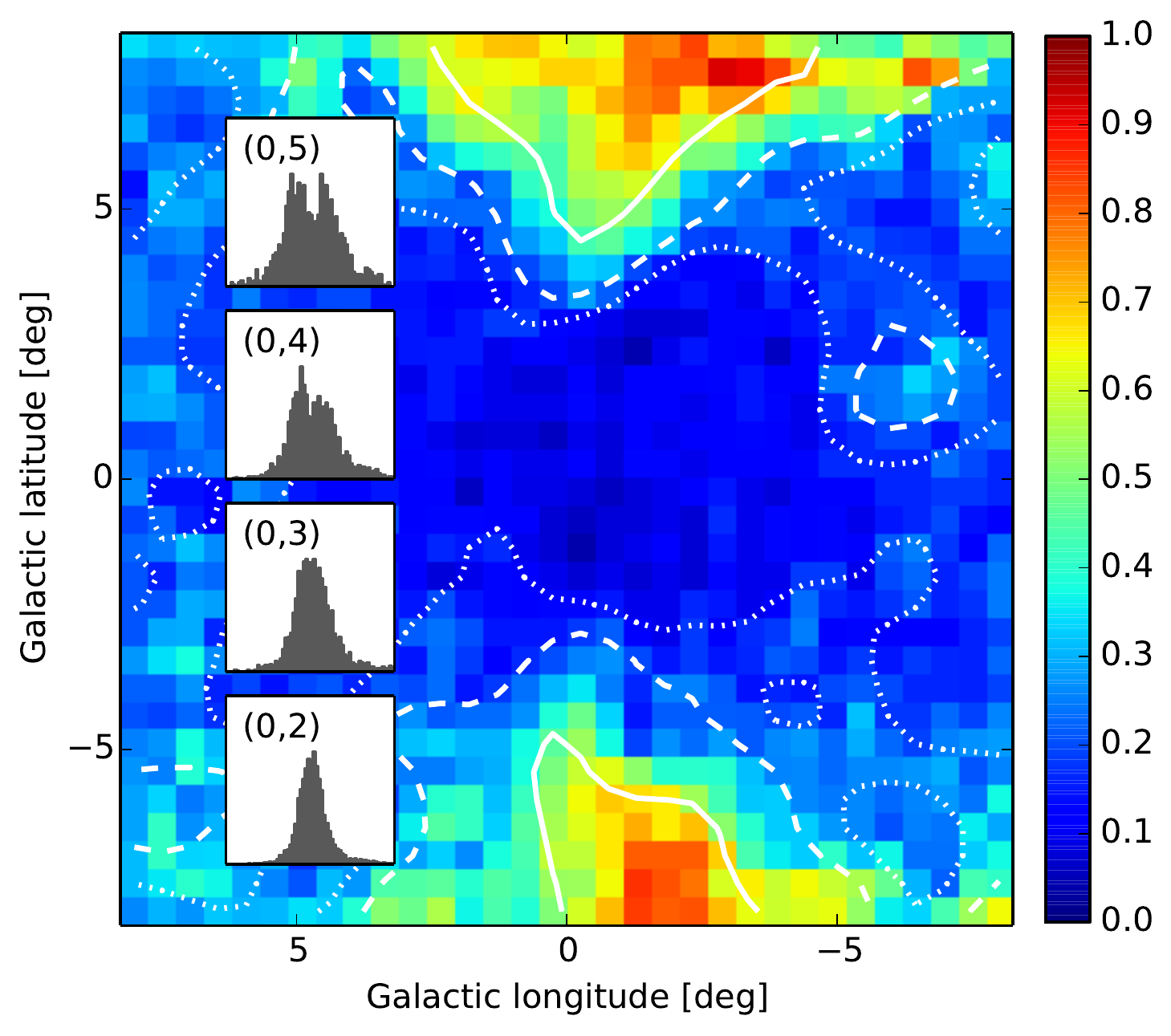}
\includegraphics[width=0.38\textwidth]{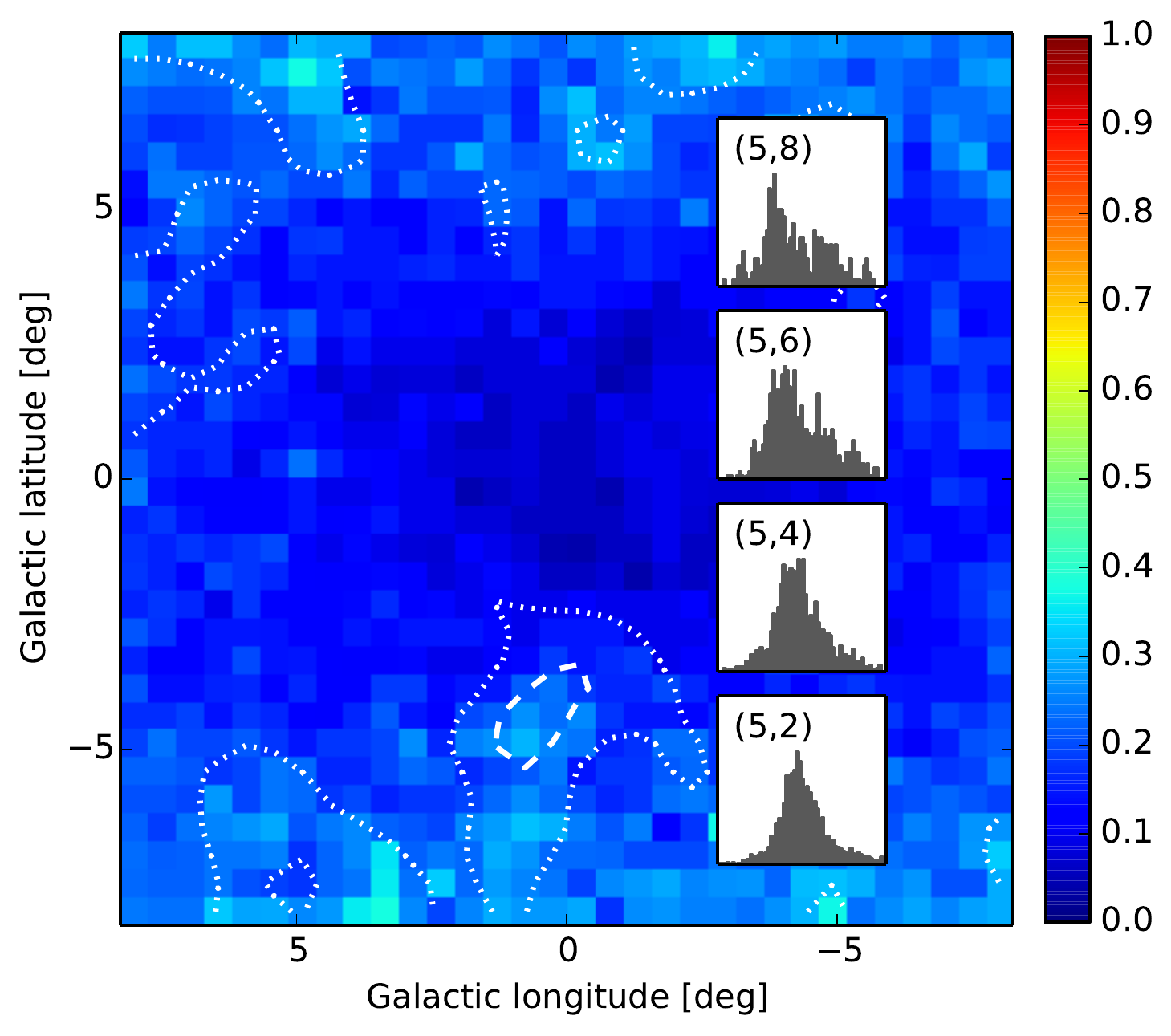}
\caption
{
Testing the double-peaked distance distribution using the kernel density estimator. The left panel shows the map of $h_1$, and right panel shows $h_2$. Units are in $\kpc$. Dotted, dashed and solid contour lines indicate the significance levels ($p_1$ and $p_2$) of $0.5$, $0.68$ and $0.95$, respectively. Histograms of distances for several typical fields are also shown in stamps with their field positions $(l,b)$. In two triangular regions above and below the disk plane ($|b|>3\degree$), we obtain large $h_1$ at high significance levels but plain $h_2$ with low $p_2$, indicating that the distance distribution is strongly bimodal there. We also find that, the critical kernel size $h_1$ increases with $|b|$ near $l\sim0\degree$, indicating that larger kernels are required to over-smooth the distribution at higher latitudes. This is consistent with the increasing peak separation with $|b|$.
}
\label{fig:silverman_test}
\end{figure*}

To further illustrate the statistical significance of the double-peaked feature in the distance distribution, the critical kernel sizes ($h_1$, $h_2$) and their significance levels ($p_1$, $p_2$) are illustrated in Figure \ref{fig:silverman_test}. Closer to the disk plane, most regions towards the Galactic bulge feature low $p_1$, $p_2$ and relatively small $h_1$, $h_2$, suggesting that the distance distributions are single-peaked there. In two triangular regions away from the disk plane ($|b|>3\degree$), we obtain clearly larger $h_1$ with high significance level, while both $h_2$ and $p_2$ are modest there. The robustly large $h_1$ but insignificant $h_2$ here indicates that in these regions, the distance distribution is strongly double-peaked due to the presence of the vertical X-shape. Furthermore, the critical kernel size $h_1$ increases with $|b|$, which also agrees with the larger peak separation at higher latitudes. Since larger kernel sizes are required to over-smooth the gap between the two modes towards higher latitudes, the increasing $h_1$ with $|b|$ also implies the vertical X-shape.

In practice, the identification of the X-shape can be affected by various uncertainties in the determination of the distance, such as sample contamination, photometric errors, non-uniform extinctions and variations in the intrinsic luminosity of RCs. Moreover, the location of the peak depends on the peak-finding method adopted. On the other hand, this model may not exactly illustrate the X-shaped feature in the MW bulge. Despite the possible disagreements between the model and observations, our model is still valuable in studying the X-shaped MW bulge for the broad similarities with the observed X-shape geometry.

\subsection{Radial Velocities}

Modern spectroscopic surveys have acquired precise measurements of stellar radial velocities, which can provide important clues to understanding the dynamics of Galactic structures. Previous radial velocity observations have revealed the cylindrical rotation of the Galactic bulge, indicating the secular nature of its formation history (Howard et al. 2009; Shen et al. 2010; Williams et al. 2011). V13 studied the radial velocity of RCs inside the X-shape towards $(0\degree, -6\degree)$. They observed an excess of approaching stars at $V_{\mathrm{los}}\sim-80\kms$ in the bright RCs (near side), and a similar excess of receding stars at $V_{\mathrm{los}}\sim80\kms$ in the faint RCs (far side). In contrast, inside the Plaut's window ($b=-8\degree$), the radial velocity barely differs between the bright and faint RCs (De Propris et al. 2011). The numerical study of G14 related the X-shape morphology to a minimum in the mean difference $V_{\mathrm{los}}$ between the two sides. Therefore, although there have been a handful of kinematic studies of the bulge stars focused in exploring the X-shape, the kinematic imprints of the X-shape are still not well studied.

\begin{figure*}[h!]
\centering
\includegraphics[width=0.35\textwidth]{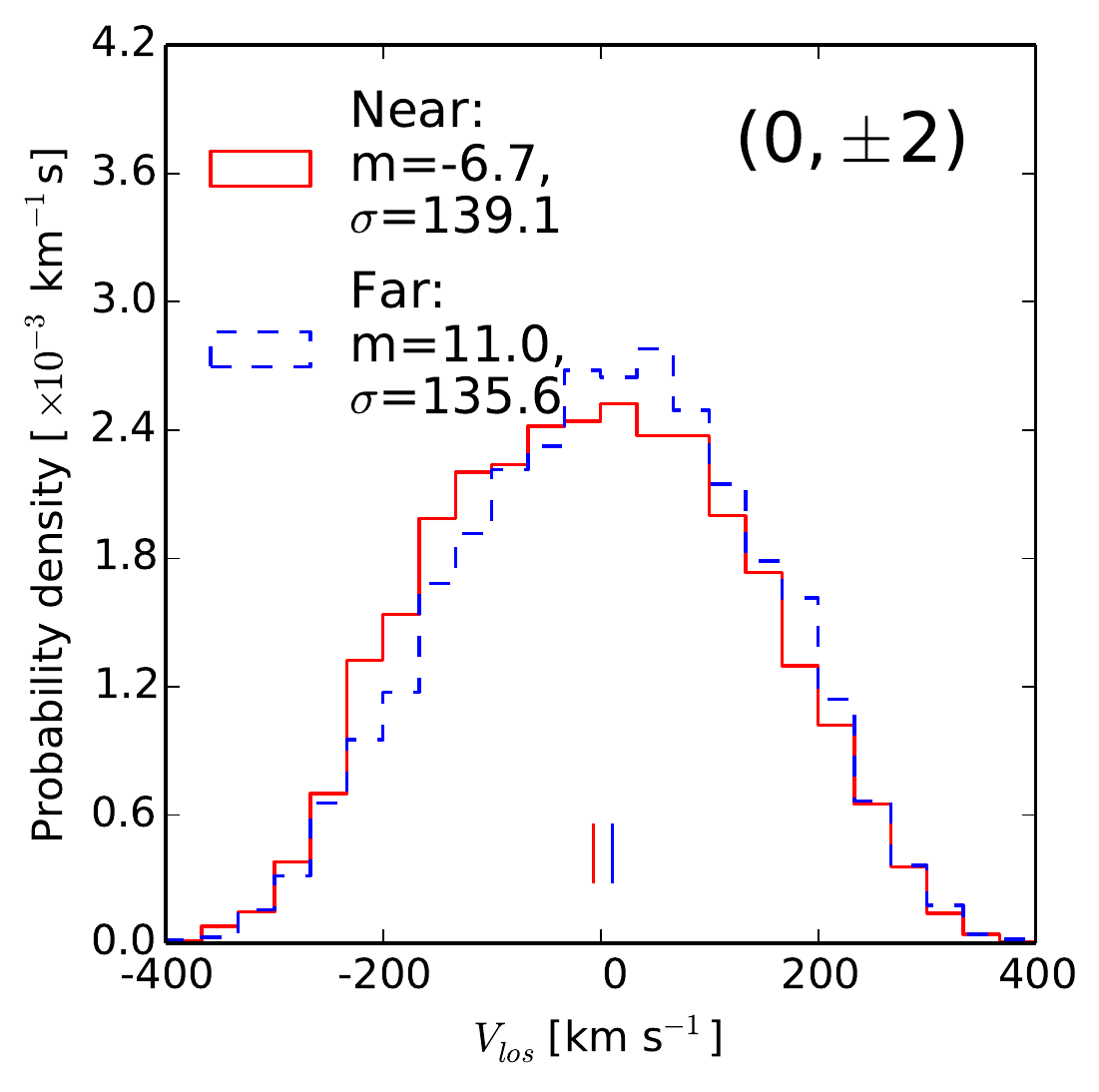}
\includegraphics[width=0.35\textwidth]{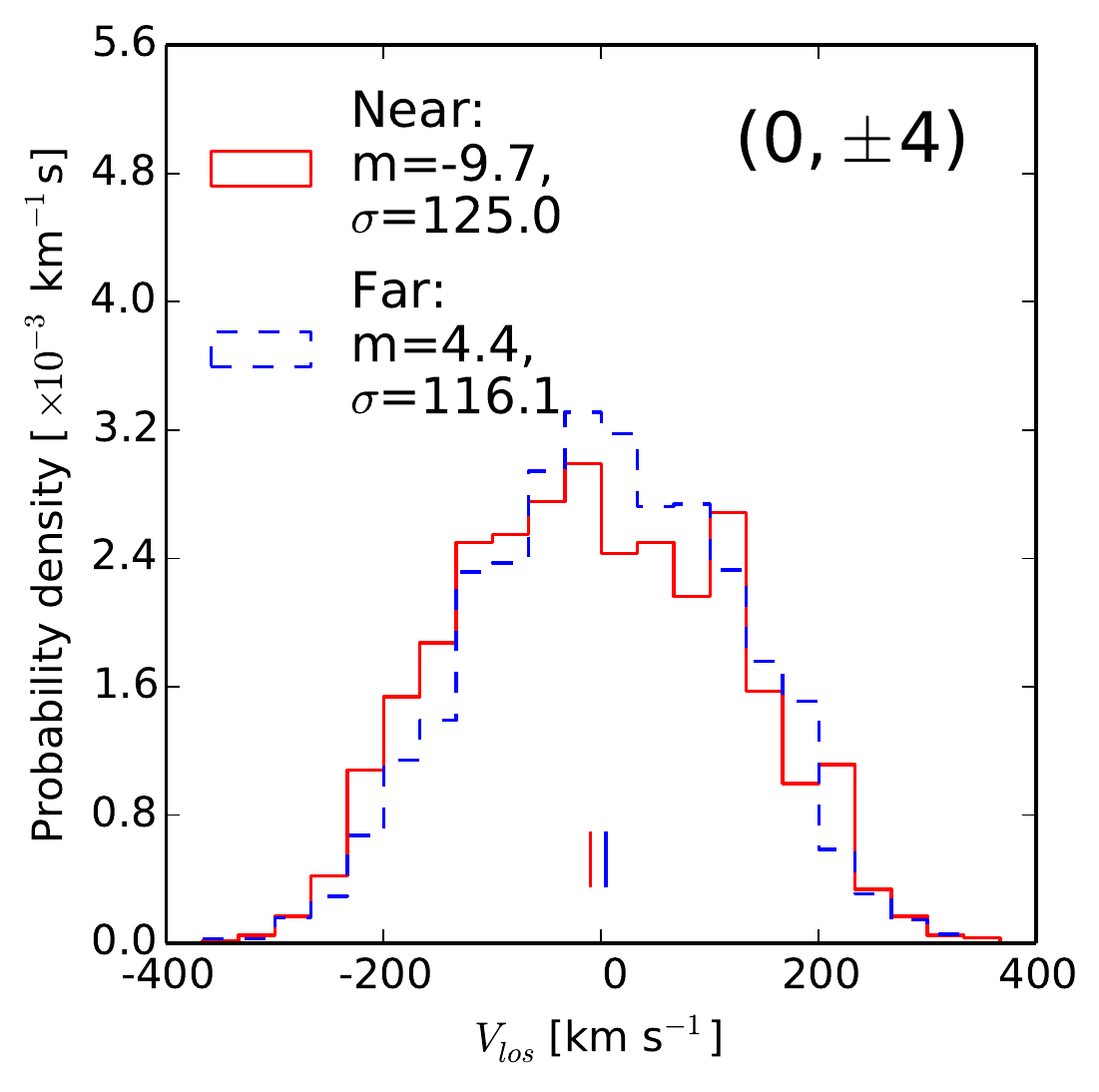}
\\
\includegraphics[width=0.35\textwidth]{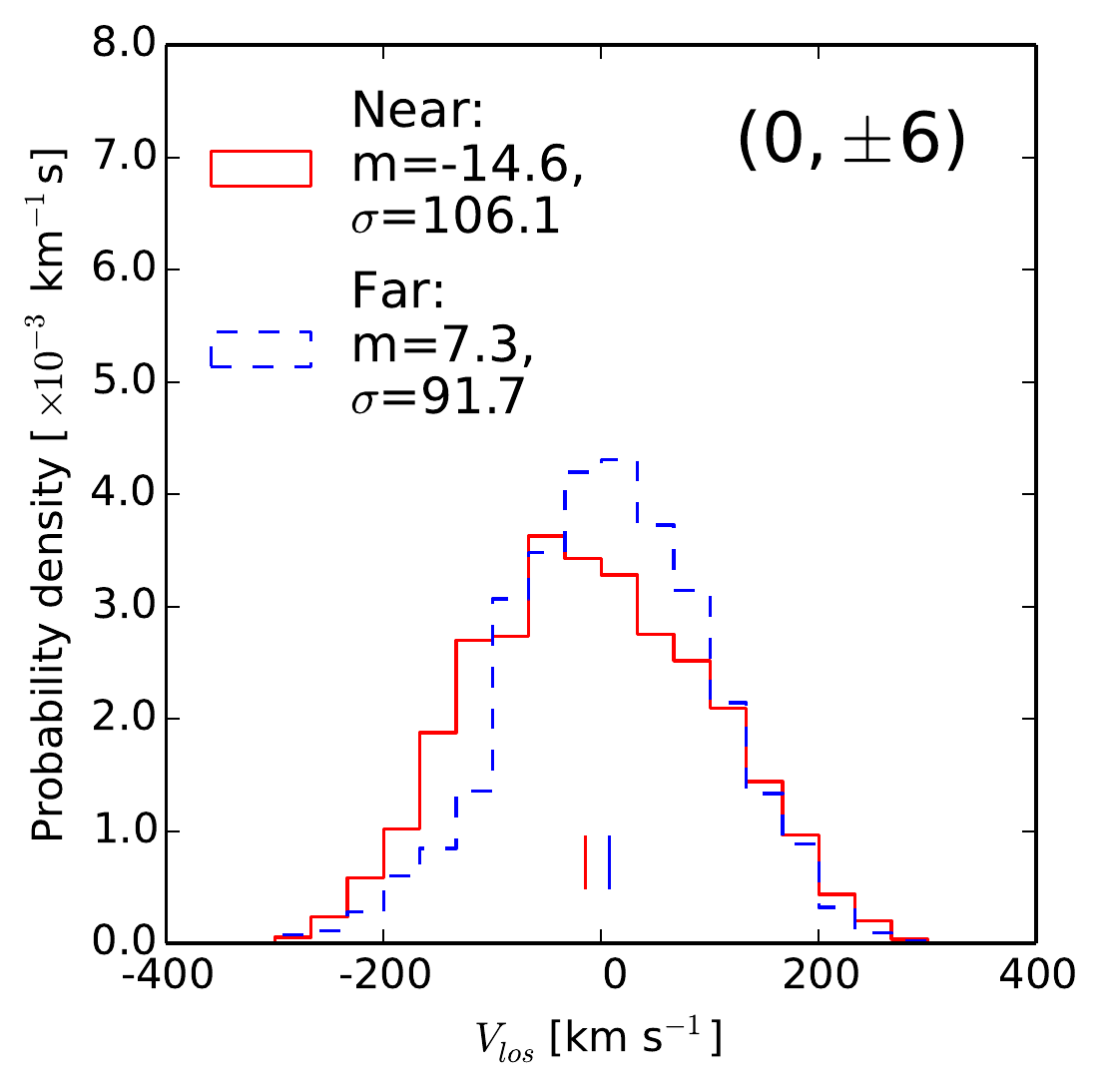}
\includegraphics[width=0.35\textwidth]{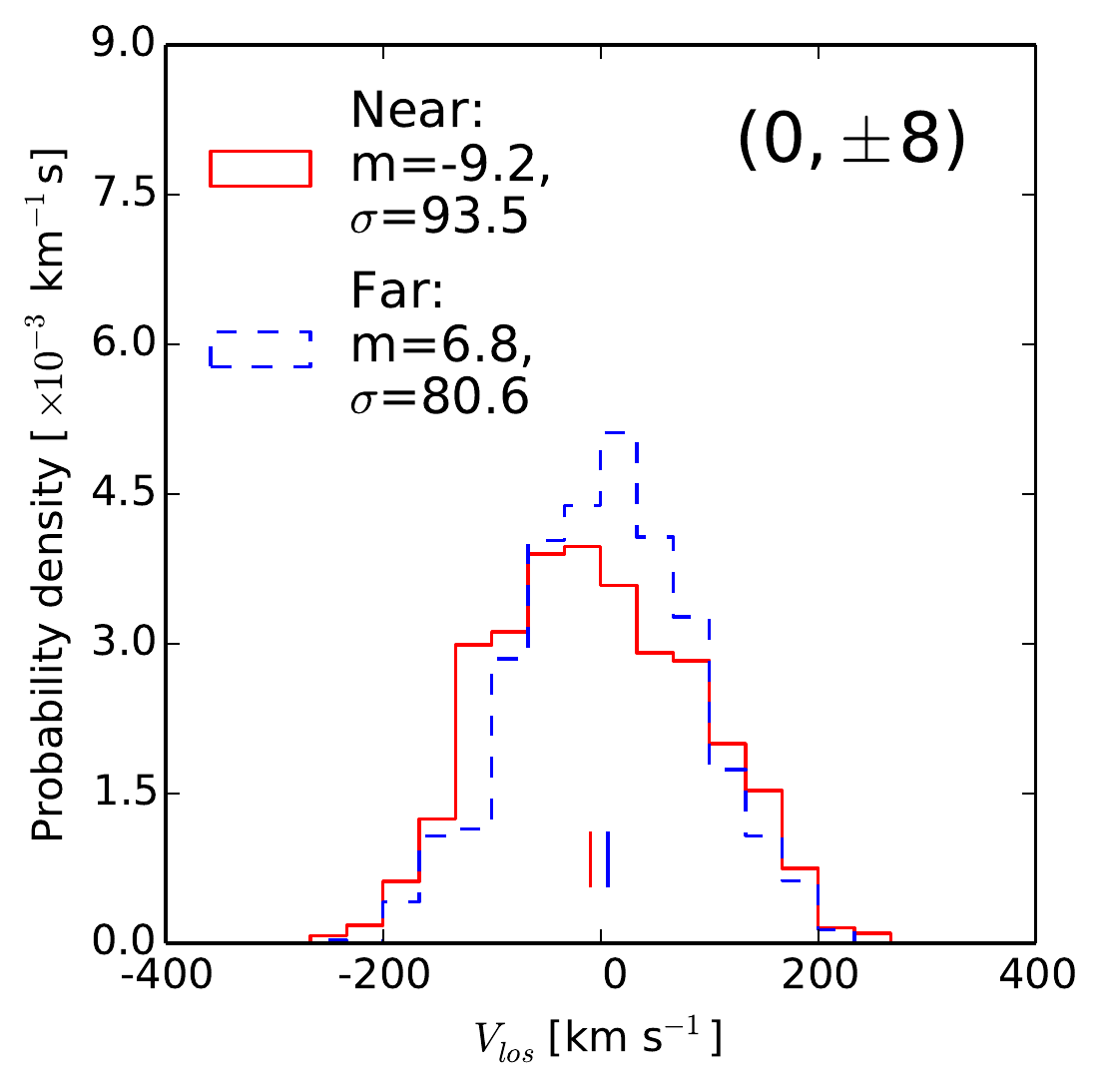}
\caption
{
Radial velocity distributions at the near and far sides in fields along $l=0\degree$, including the same field ($0\degree,-6\degree$) investigated in V13. For each panel, two $1.5\degree \times 1.5\degree$ fields with opposite $b$ are combined to obtain better number statistics, and the field position $(l,b)$ is labeled at the upper right. The near and far sides are shown in red solid lines and blue dashed lines respectively, with their median values ($\mathrm{m}$) and dispersions ($\sigma$) listed in the legend. The vertical segments indicate the median values of $V_{\mathrm{los}}$ at the two sides. We find more approaching (negative $V_{\mathrm{los}}$) particles at the near side,
and more receding (positive $V_{\mathrm{los}}$) particles at the far side, as indicated by their negative and positive median velocities. Meanwhile, the velocity dispersions at the two sides decrease with $|b|$.
}
\label{fig:v_los}
\end{figure*}

First we study the radial velocity in fields along the minor axis of the bulge. In calculating the radial velocity, we set the circular velocity of the Sun to be $V_0 = 220 \kms$, without considering the solar motion with respect to the Local Standard of Rest (LSR). Since the model is symmetric with respect to the Galactic plane, two fields with opposite latitudes are combined in the analysis to achieve better number statistics. Figure \ref{fig:v_los} shows the radial velocity distributions in $l=0\degree$, $b = \pm 2\degree, \pm 4\degree, \pm 6\degree$ and $\pm 8\degree$. For both sides of the bar/X-shape, the radial velocity spans a wider range at latitudes closer to the Galactic plane, which is seen in the decreasing velocity dispersion with $|b|$. In the same field, we find systematically smaller velocity dispersion at the far side, indicating a decreasing velocity dispersion away from the Galactic disk. In all these fields, the near side has more approaching particles (negative $V_{\mathrm{los}}$), while the far side has more receding particles (positive $V_{\mathrm{los}}$) accordingly. The excess of approaching/receding particles is evident in the median velocities of the two sides.

\begin{figure}[h!]
\centering
\includegraphics[width=\columnwidth]{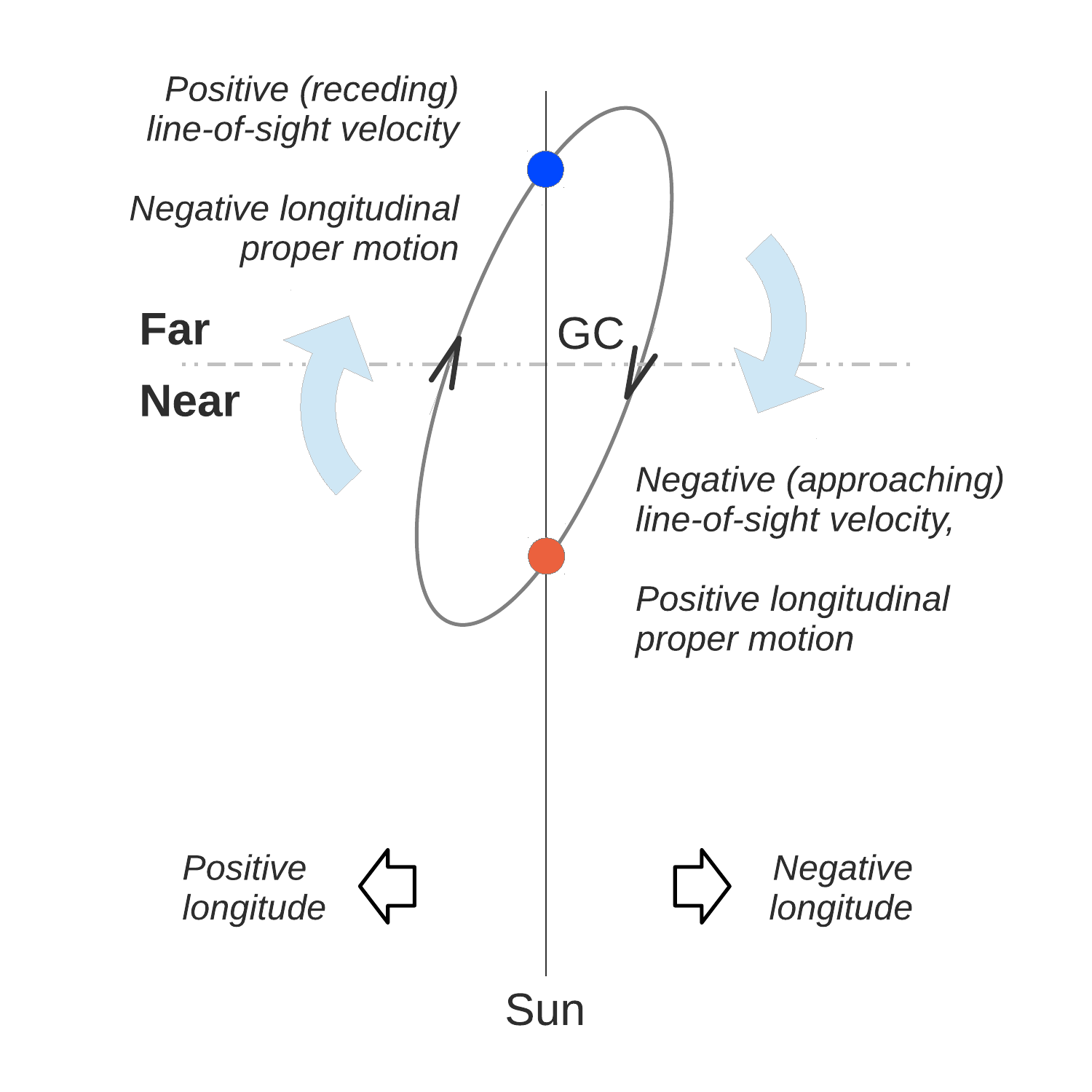}
\caption
{
The face-on sketch of a bar-supporting orbit in this configuration. Near $l=0\degree$, these orbits contribute an excess of approaching stars at the near side, and an excess of receding stars at the far side. Meanwhile, $\mu_l^{\star}$ at the near and far sides will peak at the positive and negative sides of $\mu_{l, GC}^{\star}$ respectively, where $\mu_{l, GC}^{\star}$ is the proper motion of the Galactic center due to the motion of the Sun. Its tilted alignment may also lead to the position shift in both the extrema of $\mu_l^{\star}$, and the zero-velocity line of $\overline{V}_{\mathrm{los}}$ with respect to $l=0\degree$.
}
\label{fig:bar_orbit}
\end{figure}

The excesses of $V_{\mathrm{los}}$ at the two sides may be subject to the orbital motion inside the bar structure. Figure \ref{fig:bar_orbit} sketches a naive bar-supporting orbit in this configuration. In an axisymmetric, unbarred disk, $V_{\mathrm{los}}$ at $l\sim0\degree$ is expected to be symmetric with respect to $0\kms$ at both sides. If it has coherent non-circular motions, i.e. the alignment of elongated orbits have a preferred direction, asymmetric $V_{\mathrm{los}}$ distributions with respect to $0\kms$ should be detected at both sides. In this case, an excess of approaching particles is expected at the near side, and an excess of receding particles should be observed at the far side, as expected from Figure \ref{fig:bar_orbit}. Although the reported excesses at $V_{\mathrm{los}}\sim\pm 80\kms$ in V13 are not clear in this model\footnote{We notice that at $(0\degree,-6\degree)$ alone, there are clear excesses at $V_{\mathrm{los}}\sim\pm80\kms$, which, however, become less noticeable in the combined field $(0\degree, \pm 6\degree)$. Perhaps we need better number statistics to confirm this, both in observations and models.}, there are indeed more approaching or receding particles at the two sides, which agrees well with Figure \ref{fig:bar_orbit}. Nidever et al. (2012) also used bar orbits to explain the cold high-velocity peaks seen in the commissioning observations of the Apache Point Observatory Galactic Evolution Experiment (APOGEE). However, as Li et al. (2014) demonstrated with numerical simulations, bars are not expected to display such a high velocity peak; instead, the high velocity stars are located around the tangential point between the bar orbits and the sight lines.

\begin{figure*}[h!]
\centering
\includegraphics[width=\textwidth]{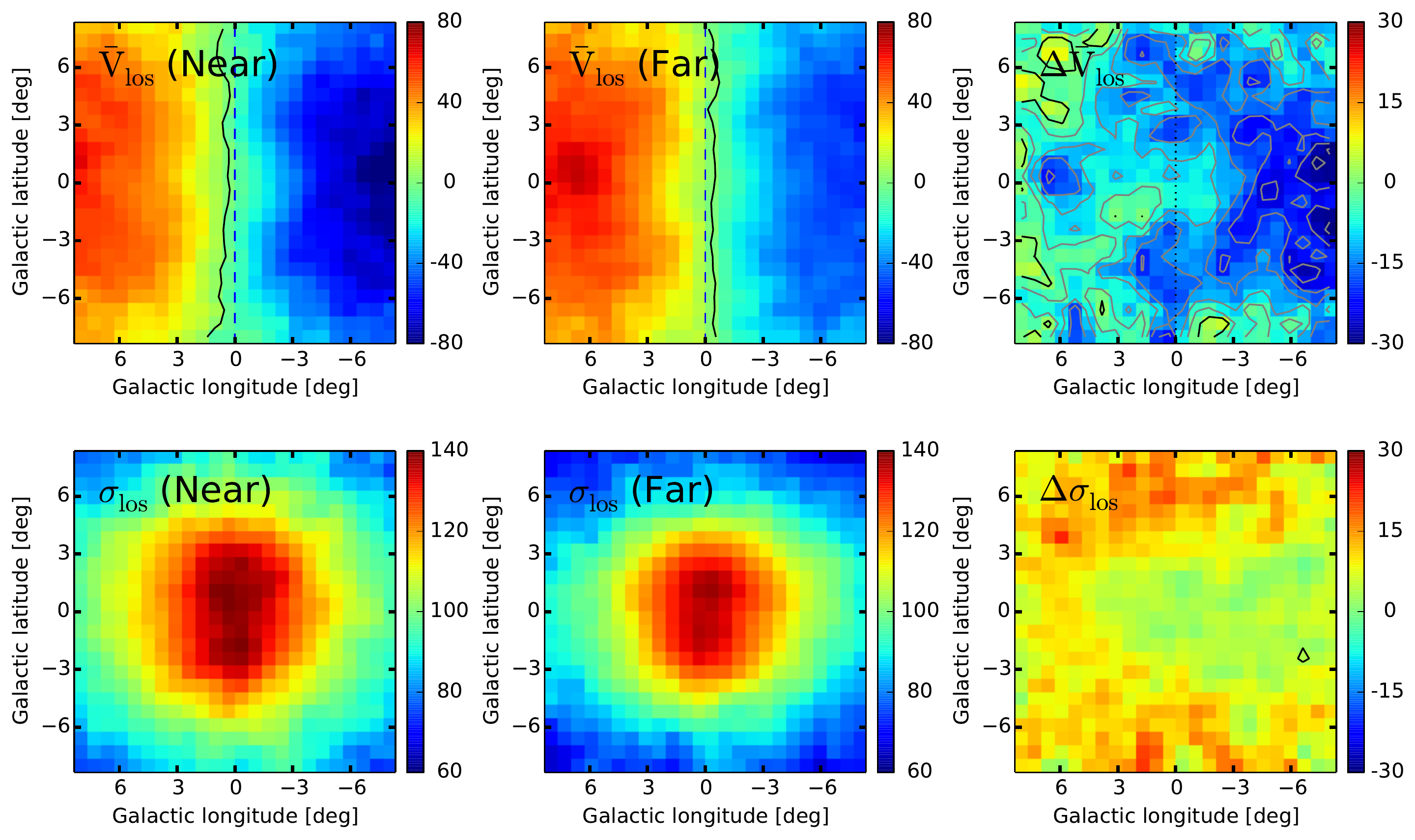}
\caption
{
The mean line-of-sight velocity ($\overline{V}_{\mathrm{los}}$) (upper) and the velocity dispersion ($\sigma_{\mathrm{los}}$) (lower) across the bulge region, in $\kms$. The left and middle columns show the near and far sides respectively, with the difference between the two sides shown in the right column (with contours). The cylindrical rotation of the bulge is seen in $\overline{V}_{\mathrm{los}}$ at both sides, and $\sigma_{\mathrm{los}}$ features a kinematically hot Galactic center. The zero-velocity line of $\overline{V}_{\mathrm{los}}$ shifts away from $l=0\degree$, which appears at $l>0\degree$ at the near side, and $l<0\degree$ at the far side. Such shift agrees with the simple bar orbits sketched in Figure \ref{fig:bar_orbit}. The minimum of $\Delta\overline{V}_{\mathrm{los}}$ along $l\sim0\degree$ described in G14 is also clearly visible in our model (upper right).
}
\label{fig:vlos_map}
\end{figure*}

Then we study the mean radial velocity ($\overline{V}_{\mathrm{los}}$), and its dispersion ($\sigma_{\mathrm{los}}$) across the whole bulge region. As illustrated in Figure \ref{fig:vlos_map}, $\overline{V}_{\mathrm{los}}$ at both sides shows clear cylindrical rotation, and the longitudinal gradient of $\overline{V}_{\mathrm{los}}$ is strong at $|b|\sim3\degree$, which coincides with the latitude where the X-shape is detectable in this model. The zero-velocity line of $\overline{V}_{\mathrm{los}}$ shifts towards $l>0\degree$ at the near side, and towards $l<0\degree$ at the far side. The shift of zero-velocity lines agrees with the coherent alignment of elongated orbits. As sketched in Figure \ref{fig:bar_orbit}, the radial velocity is zero only close to the two tips of such orbits. If the bar structure is populated by such orbits, the zero-velocity lines at the two sides of the bar will appear at two opposite sides of $l=0\degree$. Also, the decreasing radial velocity dispersion away from the GC indicates a kinematically hot inner bulge. We find that, the difference of $\overline{V}_{\mathrm{los}}$ between the near and far sides (near - far) is generally negative, while the difference of $\sigma_{\mathrm{los}}$ is mostly positive. Since on the $(l,b)$ plane, the particles at the near side are closer to the disk plane than those at the far side, the positive $\Delta\sigma_{\mathrm{los}}$ and negative $\Delta\overline{V}_{\mathrm{los}}$ may reflect the kinematic gradient in the vertical direction. Near $|b|=4\degree$, we also notice the minimum of $\Delta\overline{V}_{\mathrm{los}}$ along $l=0\degree$. Considering both the absolute value and the relative depth with respect to the GC, the minimum in our model is quite similar to G14. The agreement between our model and G14 indicates that the minimum in the mean difference of radial velocity may be a coherent feature of X-shaped bulges, which may originate from the excess of approaching/receding particles at the near and far sides.

\begin{figure*}[h!]
\centering
\includegraphics[width=0.75\textwidth]{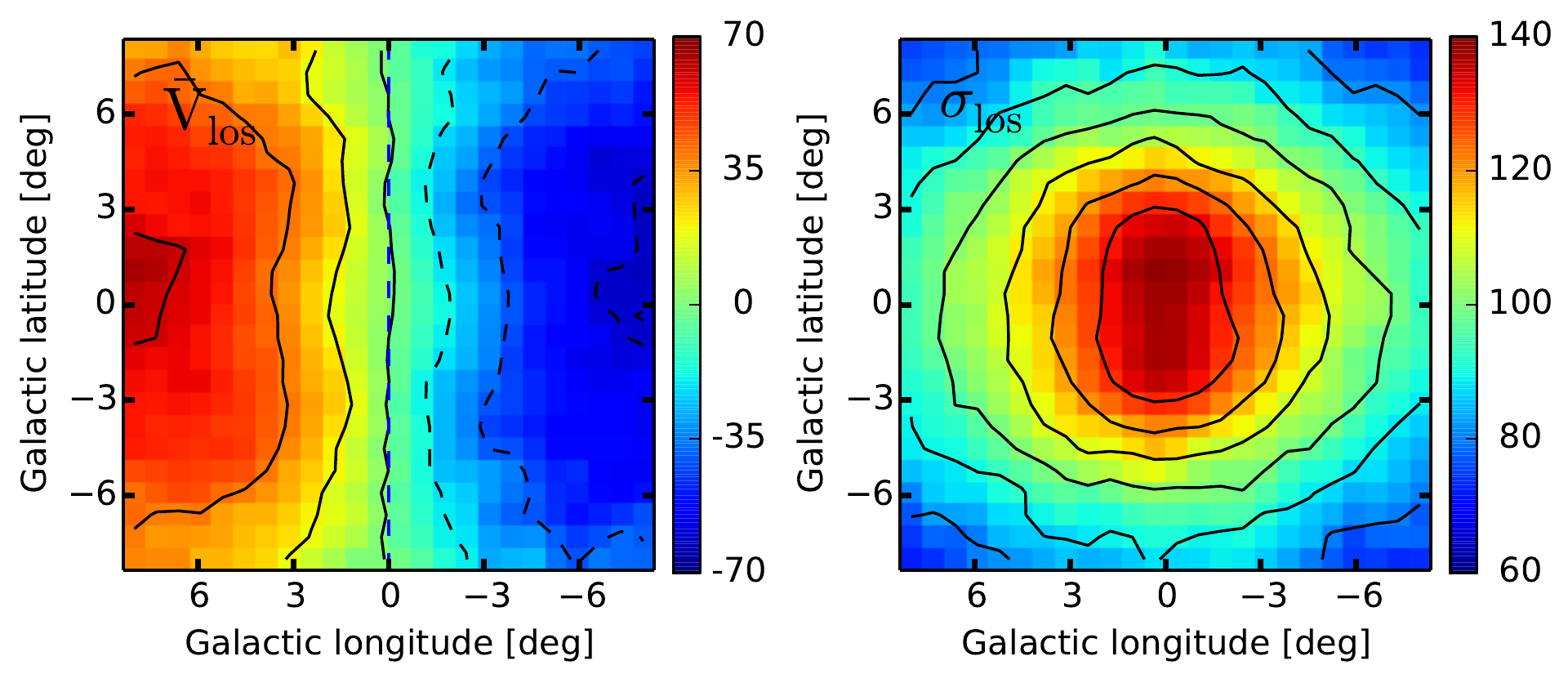}
\caption
{
The mean radial velocity (left) and the corresponding dispersion (right) across the bulge region, in $\kms$, without distinguishing the near and far sides of the galactic bar/bulge. The mean $V_{\mathrm{los}}$ (left), which clearly shows cylindrical rotation, is nearly symmetric with respect to $l=0\degree$. Dashed and solid contours show negative and positive values, respectively. The zero-velocity line has no significant deviation away from $l=0\degree$. Note that the dispersion of $V_{\mathrm{los}}$ features a vertically elongated peak near the GC, which agrees with the observed kinematic map in Zoccali et al. (2014).
}
\label{fig:v_los_all}
\end{figure*}

Zoccali et al. (2014) studied the kinematics of the Galactic bulge with the \textit{GIRAFFE} Inner Bulge Survey (GIBS). Without distinguishing the near and far sides, they mapped the line-of-sight velocity, and velocity dispersion of the bulge region using the spectra of $\sim5000$ RCs observed in 24 fields. They found an asymmetric $\overline{V}_{\mathrm{los}}$ across the Galactic longitude, and a vertically elongated peak of $\sigma_{\mathrm{los}}$ at the GC. Compared to the results in Zoccali et al. (2014), $\overline{V}_{\mathrm{los}}$ of our model is nearly symmetric with $l$ (left panel of Figure \ref{fig:v_los_all}), even in the presence of a tilted bar. We also find a vertically elongated peak of $\sigma_{\mathrm{los}}$ (right panel of Figure \ref{fig:v_los_all}). Although the velocity dispersion map in Zoccali et al. (2014) was interpolated from coarsely sampled fields, it still shares great similarity with our model. Therefore, the vertically elongated dispersion map in Zoccali et al. (2014) may be physically meaningful; it may originate from the velocity distribution of bar supporting orbits. We will study this feature in greater detail in our future work.

\subsection{Proper Motions}

The proper motion reflects the angular velocity of stars in the heliocentric frame. It is useful to study the dynamics of the MW (e.g. Koz\l owski et al. 2006; Rattenbury et al. 2007; Soto et al. 2012). Previous usage of the proper motion was limited by the data quality and the lack of accurate distance estimation. New space missions like \textit{Gaia} will obtain precise proper motion and distance measurements for an unprecedented number of stars, and the study of the Milky Way dynamics will benefit significantly from the upcoming boom of high-quality data. Currently, the potential of the proper motion is still to be fully explored, and less is known about the imprints of the fine structures (like the X-shape) on the observed proper motions. In this section, we will study the proper motion towards the bulge region in our model.

\begin{figure*}[ht!]
\centering
\includegraphics[width=0.35\textwidth]{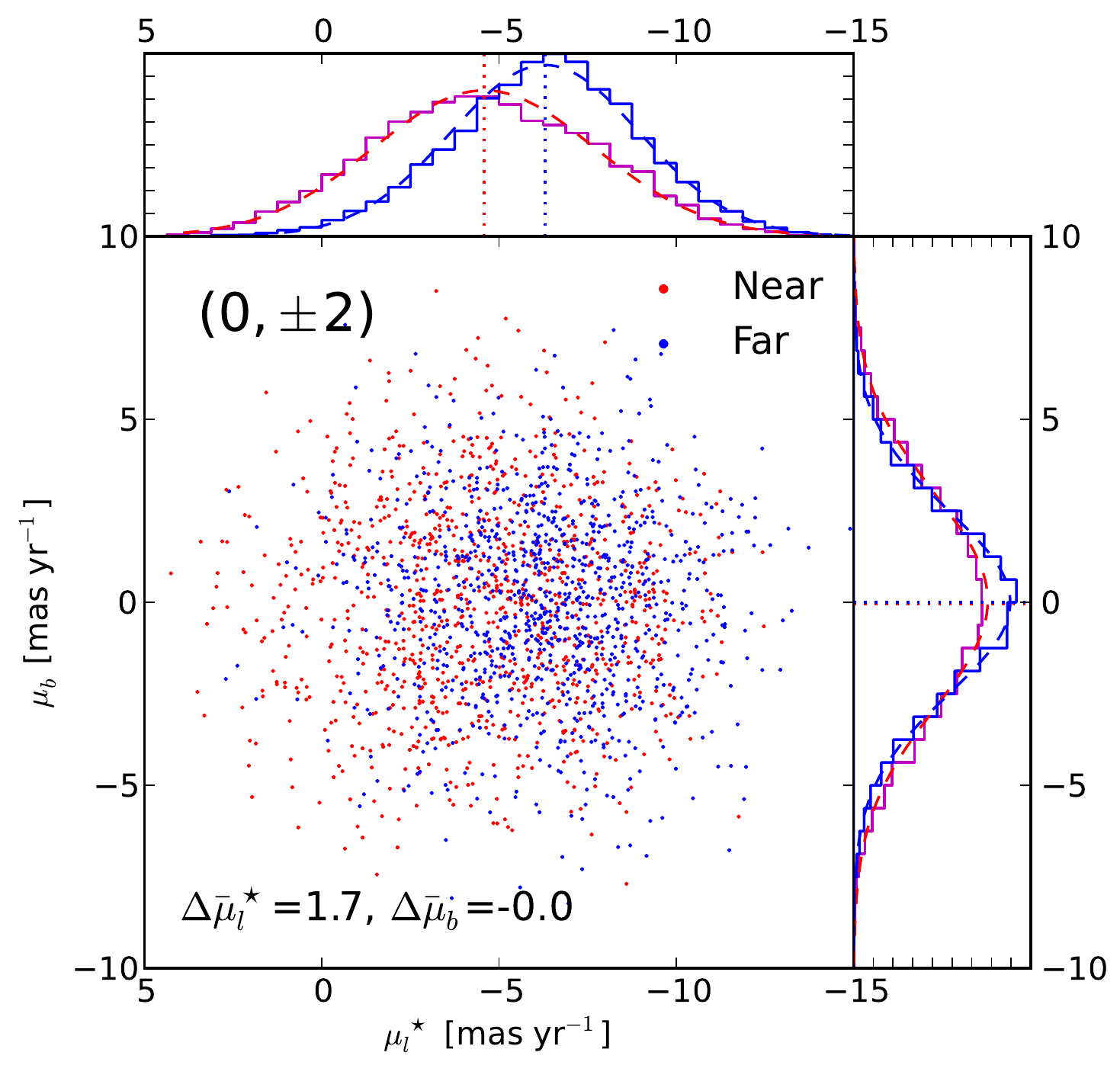}
\includegraphics[width=0.35\textwidth]{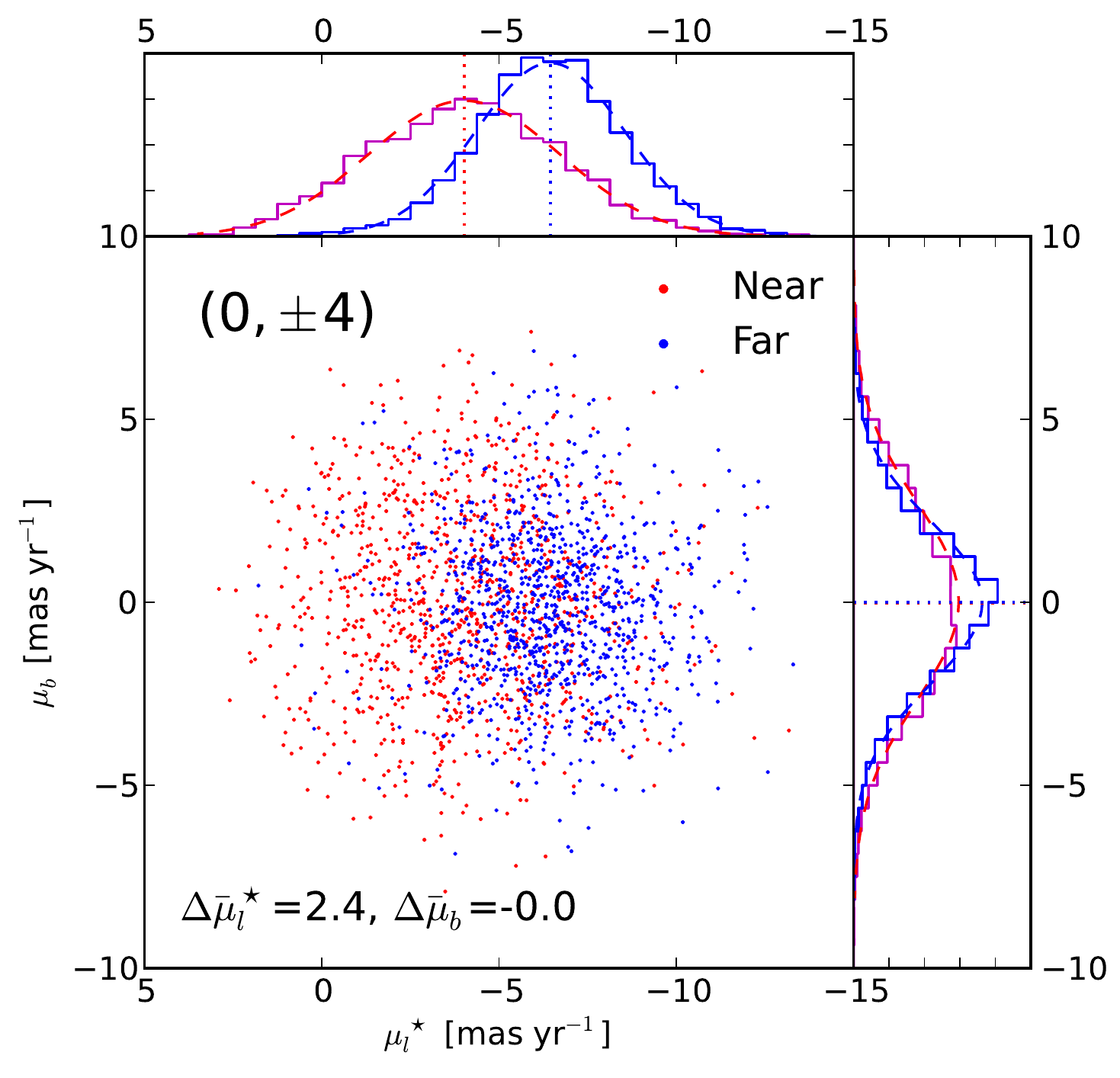}
\\
\includegraphics[width=0.35\textwidth]{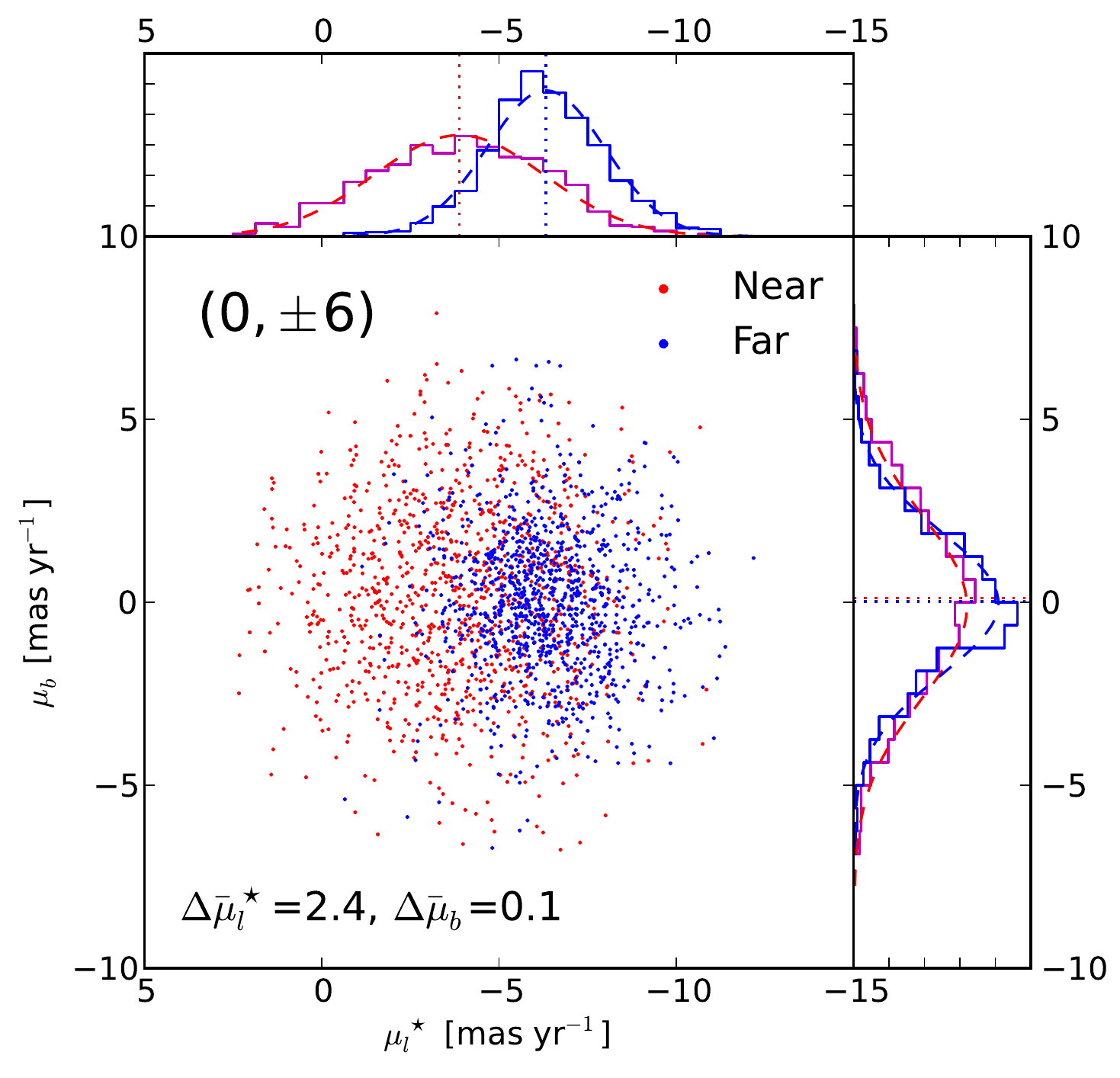}
\includegraphics[width=0.35\textwidth]{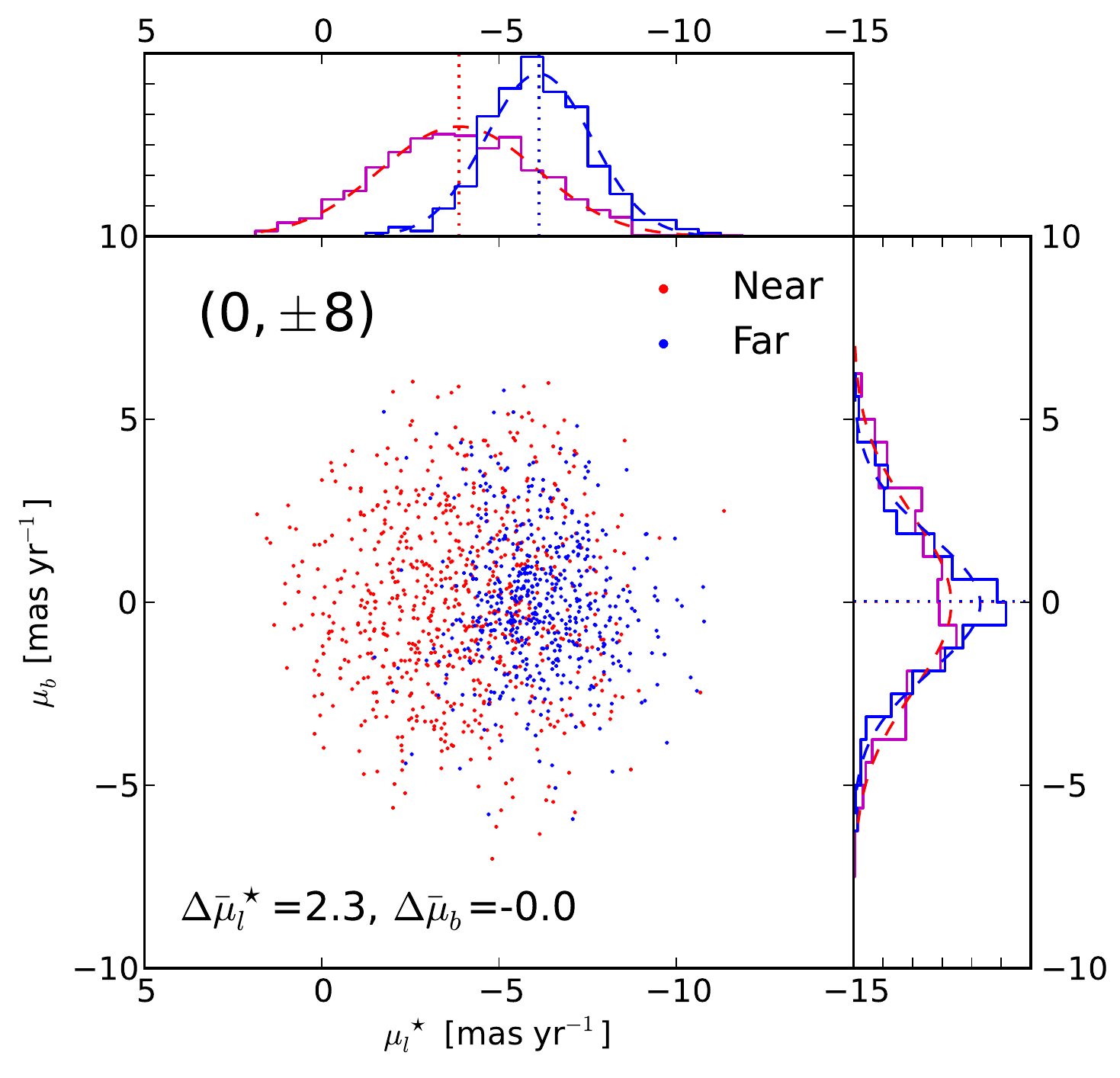}
\caption
{
Proper motions of the near and far sides at $l=0\degree$, $b=2\degree,4\degree,6\degree$ and $8\degree$. The co-added $\mu_b$ from particles at $b<0\degree$ are taken with the opposite sign as required by the symmetry of the model. Particles at the near side are in red, while the far side is shown in blue. Histograms show the distributions of $\mu_l^{\star}$ and $\mu_b$ at the two sides, with the dashed lines showing their Gaussian fit, and dotted lines marking their mean values. The $\mu_l^{\star}$ distributions peak at the positive side of $\mu_{l,GC}^{\star}$ (black solid lines) at the near side, and the negative side of $\mu_{l,GC}^{\star}$ at the far side. While for $\mu_b$, both near and far sides peak around $0\masyr$. The dispersions of $\mu_l^{\star}$ and $\mu_b$ are always larger at the near side. For both sides, all the dispersions decrease with $|b|$.
}
\label{fig:proper_motion}
\end{figure*}

First, we study the proper motion distribution along the minor axis of the bulge. Figure \ref{fig:proper_motion} shows the proper motion in four fields ($l=0\degree$, $b=2\degree,4\degree,6\degree$ and $8\degree$). To ensure robust statistics, as in the previous section, the fields with opposite latitudes are co-added to carry out the analysis. The latitudinal proper motion for fields at $b<0^\degree$ is reversed according to the symmetry of the model. For $\mu_l^{\star}$, the near and far sides peak at the positive and negative sides of $\mu_{l,GC}^{\star}$ respectively, where $\mu_{l,GC}^{\star} = V_0 / R_0 = -5.46\masyr$ is the proper motion of the GC due to the motion of the Sun (assuming $R_0=8.5\kpc$, $V_0 = 220\kms$). The displaced distributions of $\mu_l^{\star}$ at the two sides lead to a significant difference in their mean values. For $\mu_b$, the two sides have similar mean values (close to zero). We find a larger dispersion of $\mu_l^{\star}$ and $\mu_b$ at the near side, and the dispersions at both sides decrease with $|b|$. The difference of $\overline{\mu}_l^{\star}$ between the near and far sides mainly reflects the rotation of the model bulge. In these fields, $\Delta\overline{\mu}_l^{\star}$ is comparable to the dispersion of $\overline{\mu}_l^{\star}$ at both sides, which indicates the strong rotation of the model bulge. The role of bar-supporting orbits in this case still needs further investigation. In our $(0\degree,\pm6\degree)$ field we measured a relatively large $\Delta\overline{\mu}_l^{\star}$ of $2.4\masyr$, which is almost three times the result presented in V13. Such a large difference deserves further exploration and explanation.

Since the proper motion scales inversely with distance, particles at the near side will show a larger proper motion dispersion than the far side, assuming similar linear velocity distributions at the two sides. This agrees well with the larger proper motion dispersion at the near side in this model. Here we test if the larger dispersion at the near side can be explained solely with the shorter distance. We identify the density peaks in fields along the minor axis of the bulge using the method described in Section 2.1, and pick the $25\%$ particles closest to the identified peak position at each side. Then, we calculate the proper motion dispersion of these peak particles, and find the ratios of dispersions between the two sides (near to far). If the larger proper motion dispersion is mainly due to the shorter distance, the ratio of proper motion dispersion (near to far) should roughly match the ratio of peak distance (far to near). The results are summarized in Table 1. The ratio of $\sigma_l$ is marginally consistent with, but slightly larger than, the ratio of peak distance, and both of them increase with latitude. However, $\sigma_{b,Near}/\sigma_{b,Far}$ poorly matches the ratio of peak distance, and no clear trend with the latitude is observed. Therefore, the larger proper motion dispersion measured at the near side cannot be solely explained by the shorter distance. Similar to the systematically smaller line-of-sight velocity dispersion at the far side, the slightly larger $\sigma_{l,Near}/\sigma_{l,Far}$ compared to $R_{Far} / R_{Near}$ may be due to the decreasing linear velocity dispersion away from the Galactic plane.

\begin{table}
\label{tab:peak}
\begin{center}
\caption{Distance ratio and proper motion dispersion ratios between the near and far sides}
\begin{tabular}{l l l l}
\hline
Latitude    & $R_{Far} / R_{Near}$  & $\sigma_{l,Near} / \sigma_{l,Far}$  & $\sigma_{b,Near} / \sigma_{b,Far}$  \\
\hline
$4\degree$  & $1.1288 \pm 0.0095$   & $1.1426 \pm 0.0447$                 & $1.1309 \pm 0.0433$                 \\
$5\degree$  & $1.1881 \pm 0.0111$   & $1.2616 \pm 0.0609$                 & $1.2253 \pm 0.0571$                 \\
$6\degree$  & $1.2292 \pm 0.0140$   & $1.3629 \pm 0.0898$                 & $1.1006 \pm 0.0788$                 \\
$7\degree$  & $1.2782 \pm 0.0155$   & $1.3982 \pm 0.1121$                 & $1.3245 \pm 0.1248$                 \\
\hline
\end{tabular}
\end{center}
\textbf{Notes}: $R_{Far}/R_{Near}$ indicates the ratio of peak distance to the Sun (far to near). $\sigma_{l,Near} / \sigma_{l,Far}$ and $\sigma_{b,Near} / \sigma_{b,Far}$ indicate the dispersion ratio (near to far) for $\mu_l^{\star}$ and $\mu_b$. Uncertainties are derived using the bootstrap method.
\end{table}

\begin{figure*}[hb]
\centering
\includegraphics[width=\textwidth]{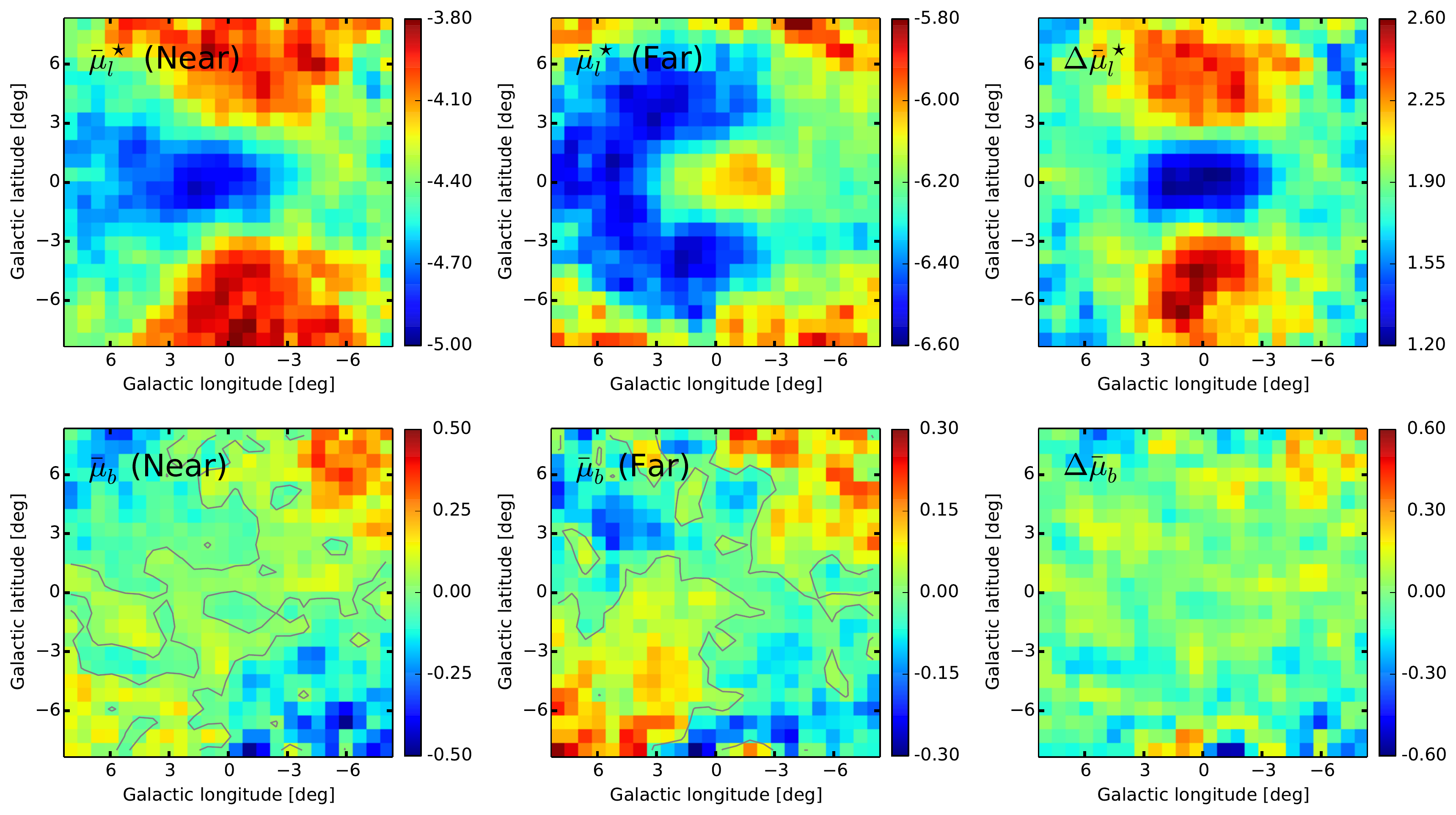}
\caption
{
The mean longitudinal proper motion (top) and the mean latitudinal proper motion (bottom) across the bulge region, in $\masyr$. The left column shows the near side, and the middle column is the far side. The mean proper motion differences between the near and far sides are shown in the right column. At each latitude, the maximum/minimum of $\overline{\mu}_l^{\star}$ shifts away from $l=0\degree$, which agrees with the coherent alignment of elongated bar orbits. Close to the Galactic plane, the maximum/minimum of $\overline{\mu}_l^{\star}$ is not prominent due to the large random motion there, and its larger displacement away from $l\sim0\degree$ is in agreement with the extended thin component of the bar. The mean latitudinal proper motion shows a clear diagonal pattern at both sides.
}
\label{fig:mean_proper_motion}
\end{figure*}

The mean proper motion and the corresponding dispersion maps across the bulge region are shown in Figure \ref{fig:mean_proper_motion}. Away from the disk plane, we find that at the same $b$, the extremum of $\overline{\mu}_l^{\star}$ shifts away from $l=0\degree$; the maximum at the near side shifts towards $l<0\degree$, while the minimum at the far side shifts towards $l>0\degree$. Close to the disk plane, in the near side, the weak maximum of $\overline{\mu}_l^{\star}$ moves further away from $l=0\degree$, and a local minimum emerges at $l>0\degree$. Meanwhile, at the far side, the minimum of $\mu_l^{\star}$ shifts further away from the GC, and a local maximum is seen at the negative $l$ side of the GC.

The behavior of $\overline{\mu}_l^{\star}$ can be boldly explained with the toy model illustrated in Figure \ref{fig:bar_orbit}. For an axisymmetric disk with cylindrical rotation, $l=0\degree$ is expected to be the maximum of $\overline{\mu}_l^{\star}$ at the near side, and the minimum of $\overline{\mu}_l^{\star}$ at the far side. Due to the coherent alignment of bar-supporting orbits, the extremum of $\overline{\mu}_l^{\star}$ will appear close to the bar's minor axis where stellar velocity is higher. This agrees with the shift of the maximum/minimum regarding $l\sim0\degree$. Close to the Galactic disk, the strong random motion would make the maximum/minimum less prominent. Compared to the higher latitudes where the inner buckled region of the bar dominates, the bar structure is longer and thinner in the disk plane. Its wider geometric extent leads to a larger displacement of the maximum/minimum close to $b\sim0\degree$, and the lower velocity at the tip of bar orbits may result in the extrema observed at the other side of the maximum/minimum.

For both near and far sides, $\overline{\mu_b}$ shows diagonal patterns across the bulge region, featuring a characteristic mean value of $\sim0.3\masyr$ (corresponds to $\sim12\kms$ at the bulge distance). This may be caused by the projection of the cylindrical rotation onto the inclined line of sight. For example, at the near side of $(6\degree,6\degree)$, $\overline{\mu}_b = -0.2\masyr$, while $\overline{V}_{\mathrm{los}} = 51\kms$. Given the mean distance of the particles ($7.6\kpc$), $\overline{\mu}_b = -0.2\masyr$ corresponds to a linear velocity of $7.2\kms$. Their ratio, $7.2\kms/51\kms = 0.141$, is close to the tangential of the latitude ($\tan 6\degree = 0.105$).

\begin{figure*}[hb!]
\centering
\includegraphics[width=0.35\textwidth]{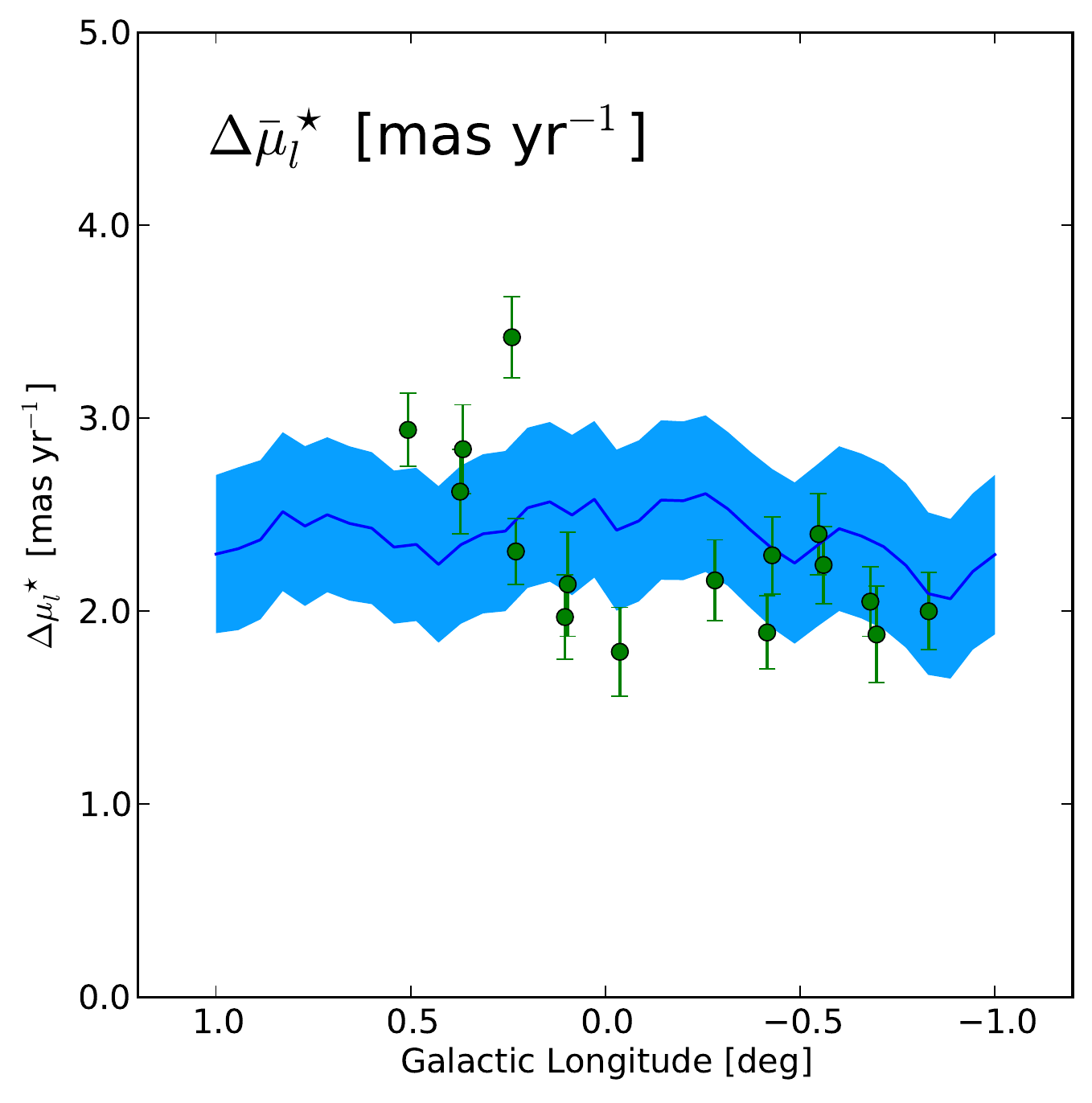}
\includegraphics[width=0.35\textwidth]{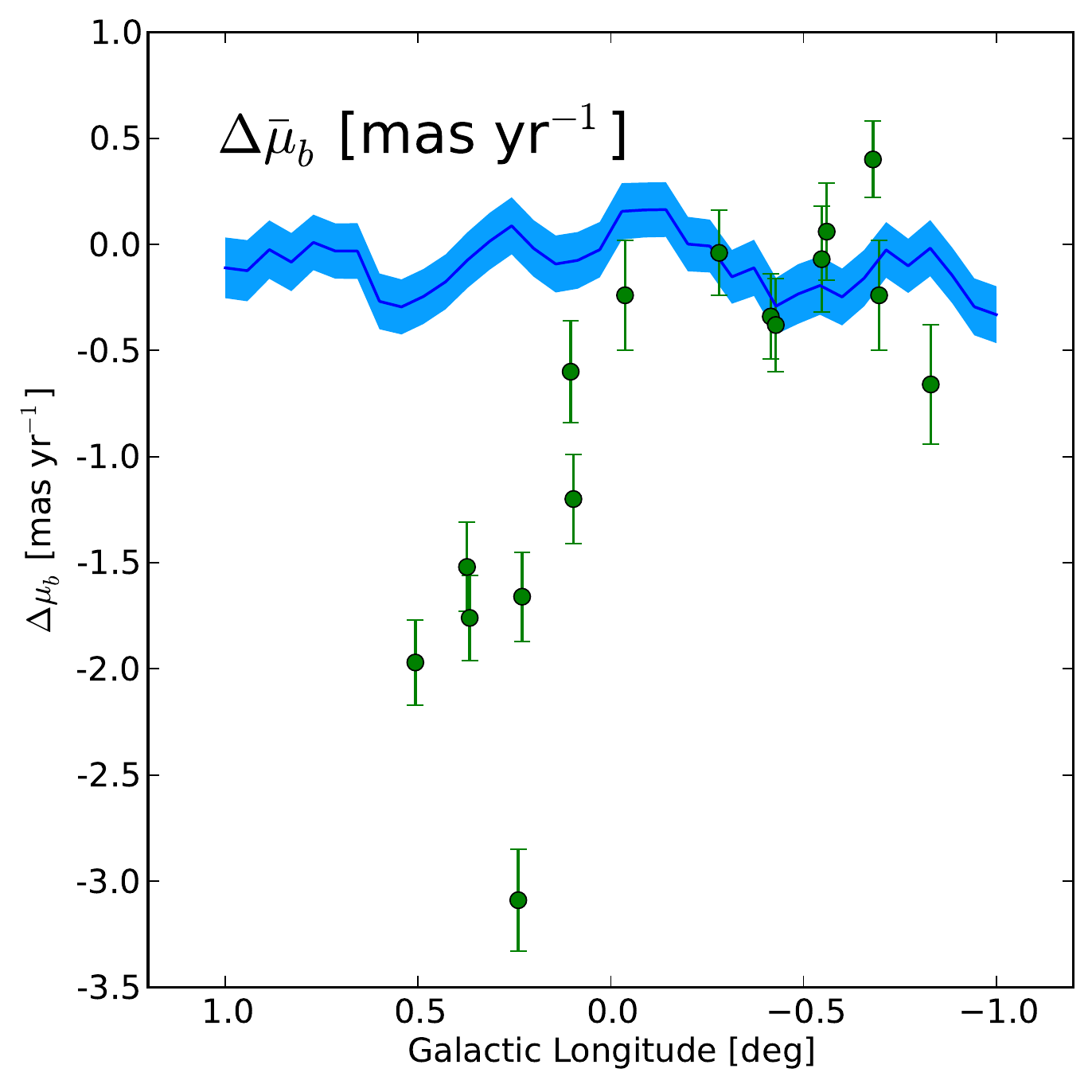}
\caption
{
The mean differences of $\mu_l^{\star}$ (left) and $\mu_b$ (right) at $b=-5\degree$, $-1\degree<l<1\degree$. In each panel, the blue lines show the mean difference, and the shaded area represents the uncertainties. Green dots with error bars are the measurements of Poleski et al. (2013) at similar positions. Inside the region, both $\Delta\overline{\mu}_l^{\star}$ and $\Delta\overline{\mu}_b$ have small variations, without any significant asymmetric feature like a break. In order to minimize the smoothing due to the large field-of-view, we reduce the field width to $0.25\degree$ here.
}
\label{fig:proper_motion_difference}
\end{figure*}

P13 studied the proper motion of RCs inside the strip $-1\degree<l<1\degree,b\sim-5\degree$. The near and far sides, identified in a probabilistic way, show an asymmetric mean difference about $l=0\degree$. At the positive longitude side where the near side of the bar dominates, $\Delta\overline{\mu}_l^{\star}$ increases with $l$ and $\Delta\overline{\mu}_b$ decreases. While at the negative longitude side, both $\Delta\overline{\mu}_l^{\star}$ and $\Delta\overline{\mu}_b$ fluctuate around constant values. The transition of $\Delta\overline{\mu}_l^{\star}$ and $\Delta\overline{\mu}_b$ was explained as the asymmetric streaming inside the X-shape. No previous model has predicted such a feature, as mentioned in P13. So we try to zoom in the same region in our model and check whether or not the X-shape can really lead to such a break of $\Delta\overline{\mu}_l^{\star}$ and $\Delta\overline{\mu}_b$ at different longitudes. Figure \ref{fig:proper_motion_difference} shows the mean difference of proper motion in the strip $-1\degree<l<1\degree, b=-5\degree$ between the near and far sides. Both $\Delta\overline{\mu}_l^{\star}$ and $\Delta\overline{\mu}_b$ are roughly symmetric with respect to $l=0\degree$; no break near $l=-0.1\degree$ is seen. Therefore, it is unclear how the bar/X-shape can really lead to the sharp break reported in P13.

\begin{figure*}[h!]
\centering
\includegraphics[width=0.65\textwidth]{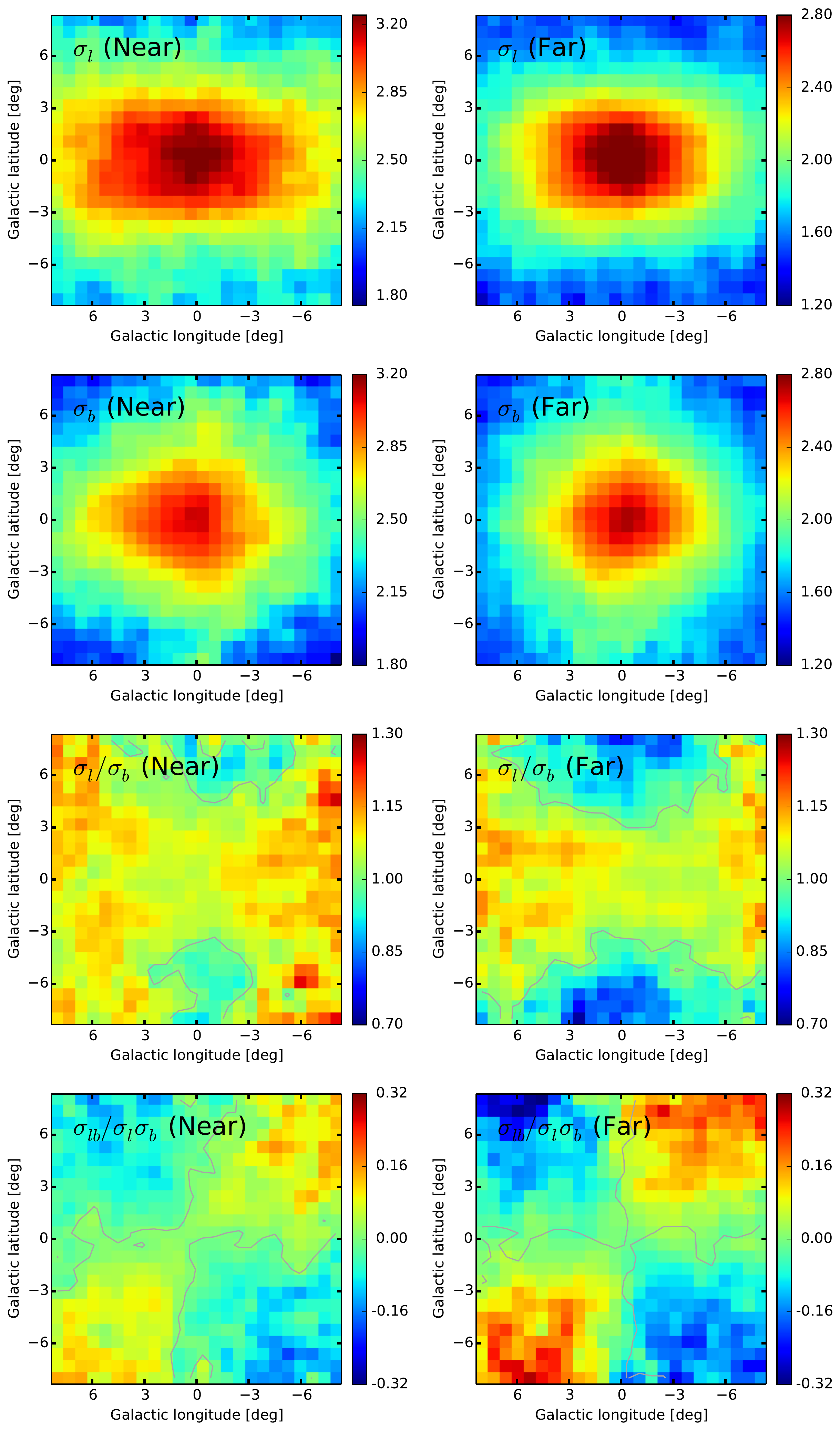}
\caption
{
The proper motion dispersion ($\sigma_l$, $\sigma_b$), the dispersion ratio ($\sigma_l / \sigma_b$) and the cross-correlation factor ($\sigma_{lb} / \sigma_l \sigma_b$) across the bulge region. The first two rows (in $\masyr$) show the proper motion dispersion ($\sigma_l$, $\sigma_b$) at the two sides. The third row represents the dispersion ratio ($\sigma_l / \sigma_b$), and gray lines mark the position where $\sigma_l / \sigma_b = 1$. The cross-correlation factor ($\sigma_{lb} / \sigma_l \sigma_b$) is at the last row, with the gray contour lines for $\sigma_{lb} / \sigma_l \sigma_b = 0$. The left and right columns correspond to the near and far sides, respectively. We find a saddle-shaped dispersion ratio, and a diagonal cross-correlation factor across the bulge region, which indicate that close to the disk plane, the direction of proper motion anisotropy tilts towards the GC, and high above the GC, the dispersion ellipse is nearly vertical.
}
\label{fig:proper_motion_dispersion}
\end{figure*}

We also map the proper motion dispersions ($\sigma_l$, $\sigma_b$) across the bulge region. The first and second rows of Figure \ref{fig:proper_motion_dispersion} show $\sigma_l$ and $\sigma_b$ measured at the two sides. In these panels, the proper motion dispersions decrease away from the GC, indicating a kinematically hot central region. The far side in our model has systematically smaller dispersions, since the particles are at larger distances than the near side. In the field analyzed in V13, $\sigma_l$ and $\sigma_b$ for both bright and faint RCs are found to be around $3\masyr$. However, our model yields smaller values in the same field, with $(\sigma_l, \sigma_b) = (2.40, 2.48)\masyr$ for the near side, and $(1.66, 1.96)\masyr$ for the far side. This is inconsistent with the result in V13 where the near and far sides share similar proper motion dispersions. In agreement with our model, the proper motion dispersions in P13 are smaller at the far side, although there are large field-to-field variations.

The distribution of proper motion in $(\mu_l^{\star}, \mu_b)$ plane partly represents the velocity anisotropy in a certain sample. We adopt two indicators to illustrate the proper motion anisotropy in this model, namely the dispersion ratio, and the cross-correlation factor. The dispersion ratio ($\sigma_l / \sigma_b$) of a sample compares the random motions between the longitudinal and latitudinal directions. In this model, the dispersion ratio is saddle-shaped at both sides (the third row in Figure \ref{fig:proper_motion_dispersion}). It is larger than unity close to the disk plane, indicating stronger random motions in the longitudinal direction. In two regions above/below the GC, the measured ratio is smaller than $1$, implying that the random motion may be stronger in the vertical direction. The proper motion distribution in V13 is less anisotropic compared to our model. Both the bright and faint RCs in V13 have dispersion ratios very close to $1$, whereas we measure $0.92$ and $0.84$ for the near and far sides, respectively. The sub-fields of P13 are spatially concentrated, but the results have large field-to-field variations, making it difficult to directly compare our model with the results.

The cross-correlation factor ($\sigma_{lb} / \sigma_l \sigma_b$, the same as the Pearson product-moment correlation coefficient $r$) measures the degree of anisotropy in the $(\mu_l^{\star}, \mu_b)$ plane (e.g. Rattenbury et al. 2007). It ranges from $0$ (no correlation) to $\pm1$ (perfect linear correlation), where positive and negative signs represent correlation and anti-correlation, respectively. However, it is worth pointing out that the factor here is insensitive to the case that the direction of anisotropy aligns with the coordinate axes. In this model, the cross-correlation factor features diagonal patterns (bottom two panels of Figure \ref{fig:proper_motion_dispersion}). Since the dispersion ratio is mostly larger than $1$ near the disk plane, the diagonal pattern implies that the direction of anisotropy tilts towards (but may not exactly pointing to) the GC. Towards higher latitudes ($|b|>3\degree$) along $l\sim0\degree$, the dispersion ratio is smaller than $1$, and the cross-correlation factor is close to zero. Therefore, the direction of proper motion anisotropy is nearly vertical there.

We also notice that at the far side, the area where $\sigma_l / \sigma_b > 1$ is larger than the near side. Meanwhile, the cross-correlation factor tends to be larger at the far side. Since the particles at the far side have greater heights than those in the near side, such results indicate that the velocity anisotropy may be stronger at larger heights. However, the analysis here lacks information on the distance and the line-of-sight velocity, so we cannot directly illustrate the three-dimensional velocity anisotropy inside the bar region. We will study the velocity anisotropy of this model and its observational imprints in our future work.

\begin{figure*}[ht!]
\centering
\includegraphics[width=0.8\textwidth]{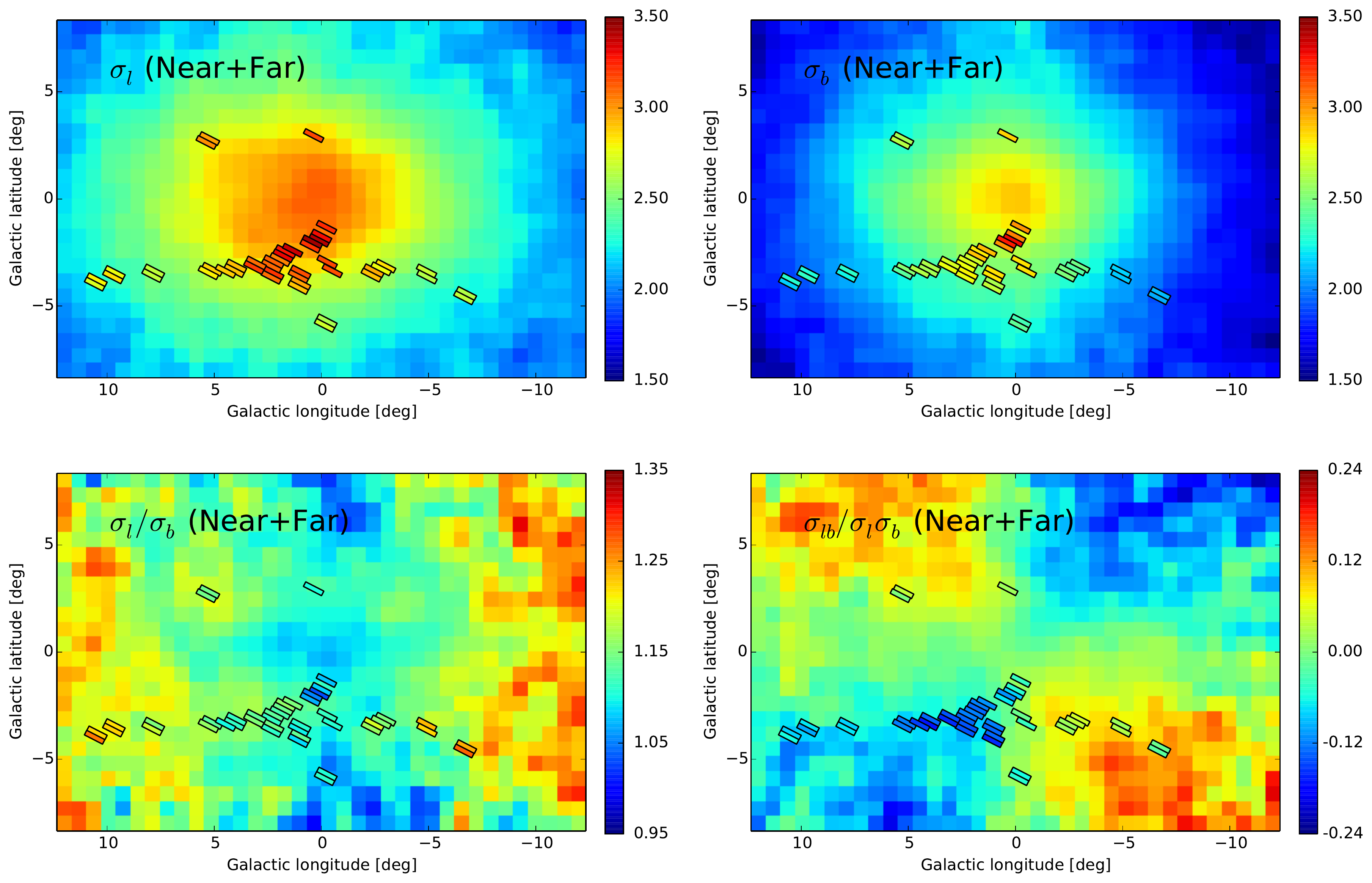}
\caption
{
Proper motion dispersions across the bulge region, without distinguishing near and far sides. The upper panels represent $\sigma_l$ and $\sigma_b$ towards the bulge region (in $\masyr$), along with the dispersion ratio $\sigma_l / \sigma_b$ and the cross correlation factor $\sigma_{lb} / \sigma_l \sigma_b$ shown in the lower left and lower right panels respectively. In each panel, the results of Rattenbury et al. (2007) are marked with boxes. Compared to the same fields in Rattenbury et al. (2007), our model has slightly lower velocity dispersions ($\sim0.3\masyr$ on average), but the dispersion ratio is very similar (the observed value is on average $0.3\pm0.6\%$ higher than the model value). For the cross-correlation factor, considering the degree of anisotropy only (taking its absolute value), we find on average $2.61\pm0.58$ times higher value in our model compared to the observed one in Rattenbury et al. (2007).
}
\label{fig:proper_motion_dispersion_all}
\end{figure*}

Rattenbury et al. (2007) mapped the proper motion dispersion across the bulge region without distinguishing the two sides of the bar. Compared to their results, the proper motion dispersions are $\sim0.3\masyr$ smaller in this model (upper two panels in Figure \ref{fig:proper_motion_dispersion_all}). For the proper motion dispersion ratio, our model well matches the observed results in Rattenbury et al. (2007). If we ignore the direction of the correlation, i.e. the sign of $\sigma_{lb}/\sigma_l \sigma_b$, and compare the degree of anisotropy using its absolute value only, our model is slightly more anisotropic compared to Rattenbury et al. (2007), as shown in the lower right panel of Figure \ref{fig:proper_motion_dispersion_all}.

\begin{figure*}[hb!]
\centering
\includegraphics[width=0.8\textwidth]{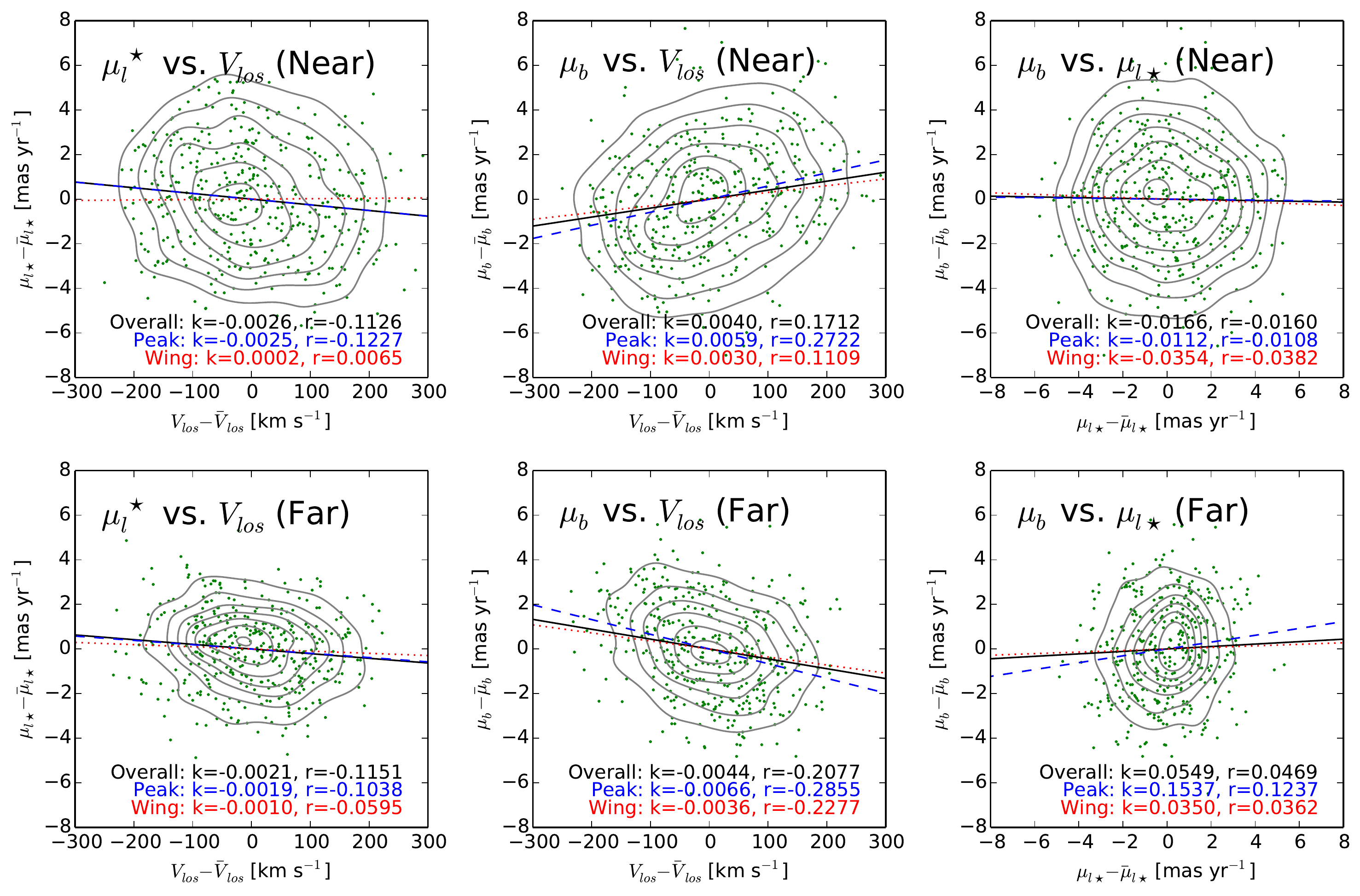}
\caption
{
Correlation of the proper motion ($\mu_l^{\star}$, $\mu_b$) and radial velocity $V_{\mathrm{los}}$ at $(0\degree,\pm6\degree)$. In each panel the contours show the density of all the particles within the near and far sides (``overall''). Green dots mark one quarter of particles closest to the peak identified at the given side (``peak''). Linear regressions of the overall sample, peak particles, and one quarter of particles furthest from the GC (``wing'') from the peaks are shown in black solid lines, blue dashed lines and red dotted lines, respectively. The slopes and Pearson's r-values are shown with the legend in each panel. For $\mu_b$ vs. $V_{\mathrm{los}}$, the slope and Pearson's r are largest in the peak sample. For $\mu_b$ vs. $\mu_l^{\star}$, at the far side, the peak sample also has largest slope and Pearson's r, while at the near side, the peak sample has smallest slope and Pearson's r. 
}
\label{fig:velocity_correlation}
\end{figure*}

\subsection{Three-dimensional Kinematics inside the X-shape}

To explore the origin and properties of the X-shaped feature in the MW, one may need to study the three-dimensional velocity distribution inside the X-shape and then distinguish the possible orbital families that result in this feature. Previous studies suggested that the boxy/peanut bulge and the X-shape correspond to the banana orbits, i.e. x1 orbits with the 2:1 vertical resonance (e.g. Pfenniger \& Friedli 1991; Patsis et al. 2002; Athanassoula 2005), but other orbital families were also suggested to support the X-shape (e.g. Portail et al. 2015b). In this scenario, we would expect stronger coherent motions inside the X-shape due to the clustering of certain resonant orbits. V13 identified the streaming motion along the bar with the correlation of $V_{los}$ vs. $\mu_l^{\star}$, but they did not find any significant correlation in $V_{los}$ vs. $\mu_b$. Here we study the stellar velocity inside the X-shape using a similar approach, and examine if the X-shape possesses unique kinematic properties.

We choose $(0\degree,6\degree)$ as our target field, and the opposite field $(0\degree,-6\degree)$ (the V13 field) was added (with negative $\mu_b$) to increase number statistics. This field is chosen because the X-shape is well-identified there, and the sufficient density of particles inside this field ensures robust statistics. For each side, among all the particles (``overall''), one quarter of particles closest to the identified peak are picked as the ``peak'' sample, while ``wing'' sample contains one quarter of particles furthest away from the GC. Rather than using a distance cut, the sample selection is based on the fraction of particles away from the peak position, since the ``thickness'' of the X-shape is not well-defined. For each sample, we use the Pearson product-moment correlation coefficient (Pearson's r) to indicate the significance of the correlation. We also perform linear regressions to find the best-fitting slopes of these samples, since such slopes may be weakened in the presence of uncorrelated data, which can then be used to compare the peak particles to the rest samples.

The results are shown in Figure \ref{fig:velocity_correlation}. We find that the peak samples (green dots) have similar proper motion and radial velocity distributions with the overall samples (contours). In other words, the peak samples do not contain or represent any distinct kinematic sub-groups. However, for $\mu_b$ vs. $\mu_l^{\star}$ and $\mu_b$ vs. $V_{\mathrm{los}}$, the slopes and Pearson's r are larger for the peak sample, and smaller for the wing sample, except for $\mu_b$ vs. $\mu_l^{\star}$ at the near side, where the peak sample has the smallest slope and Pearson's r. For $\mu_l^{\star}$ vs. $V_{\mathrm{los}}$, we do not find similar trends. The anti-correlation between $\mu_l^{\star}$ and $V_{\mathrm{los}}$ seen in the model is in agreement with V13, although they did not compare the ``peak sample'' to other stars in the same field. However, the correlation of $\mu_b$ vs. $V_{\mathrm{los}}$, which is reported to be trivial in V13, is of similar significance as $\mu_l^{\star}$ vs. $V_{\mathrm{los}}$ in this model.

The possibly stronger correlations in the peak samples indicate that the stellar velocity might be slightly more anisotropic inside the X-shape. If the X-shape is an ensemble of certain resonant orbits, the slightly stronger anisotropy inside it seems to be a sensible expectation. We notice that the correlations here may match the case of downward-tilted banana orbits (cap-like ones from the edge-on view). The downward-tilted banana orbits have their abdomens and two tips above and below the Galactic plane, respectively. Close to their abdomens, the positions where particles move downwards can explain the observed correlations. Given the clockwise rotation and the bar angle in this model, at the near side, a positive correlation between $V_{\mathrm{los}}$ and $\mu_l^{\star}$, a positive correlation between $V_{\mathrm{los}}$ and $\mu_b$ and a negative correlation between $\mu_l^{\star}$ and $\mu_b$ are expected; while at the far side, there should be a negative correlation between $V_{\mathrm{los}}$ and $\mu_l^{\star}$, a negative correlation between $V_{\mathrm{los}}$ and $\mu_b$ and a positive correlation between $\mu_l^{\star}$ and $\mu_b$. The results in this model qualitatively match the expected correlations. Moreover, $\mu_l^{\star}$ vs. $V_{\mathrm{los}}$ mainly reflects the coherent alignment of bar orbits; it is less relevant to the vertical motions. Therefore, we may not see stronger correlations of $\mu_l^{\star}$ vs. $V_{\mathrm{los}}$ in the peak samples.

However, the explanation above may be inconclusive. The correlations due to the downward-tilted, if exist, may be seriously diluted by upward-tilted banana orbits, and other orbits in the bar region. Moreover, the X-shape itself may be dominated by other orbital families (e.g. Portail 2015b), whose kinematic behavior is still poorly understood. More information is required to clarify the role of different types of orbits in the composition of the X-shape, such as their spatial extent, fraction inside the bar/bulge region, energy distribution, etc. It is worth pointing out that if the kinematics of the X-shape is dominated by the streaming motions due to banana orbits, different distributions are expected between the peak particles and the overall samples. However, the peak particles in Figure \ref{fig:velocity_correlation} show similar extended velocity distributions as the overall samples. On the other hand, if the banana orbits indeed make the X-shape, the density peaks we find along the line of sight may be in the region where particles move slower and stay longer, and thus diluting the velocity anisotropy. Therefore, although we find possible signatures of banana orbits from linear regressions, it is still too early to conclude that the banana orbits eventually lead to the observed correlations.

\subsection{Proper Motion as a Probe of Bulge Rotation}

\begin{figure*}[ht!]
\centering
\includegraphics[width=0.85\textwidth]{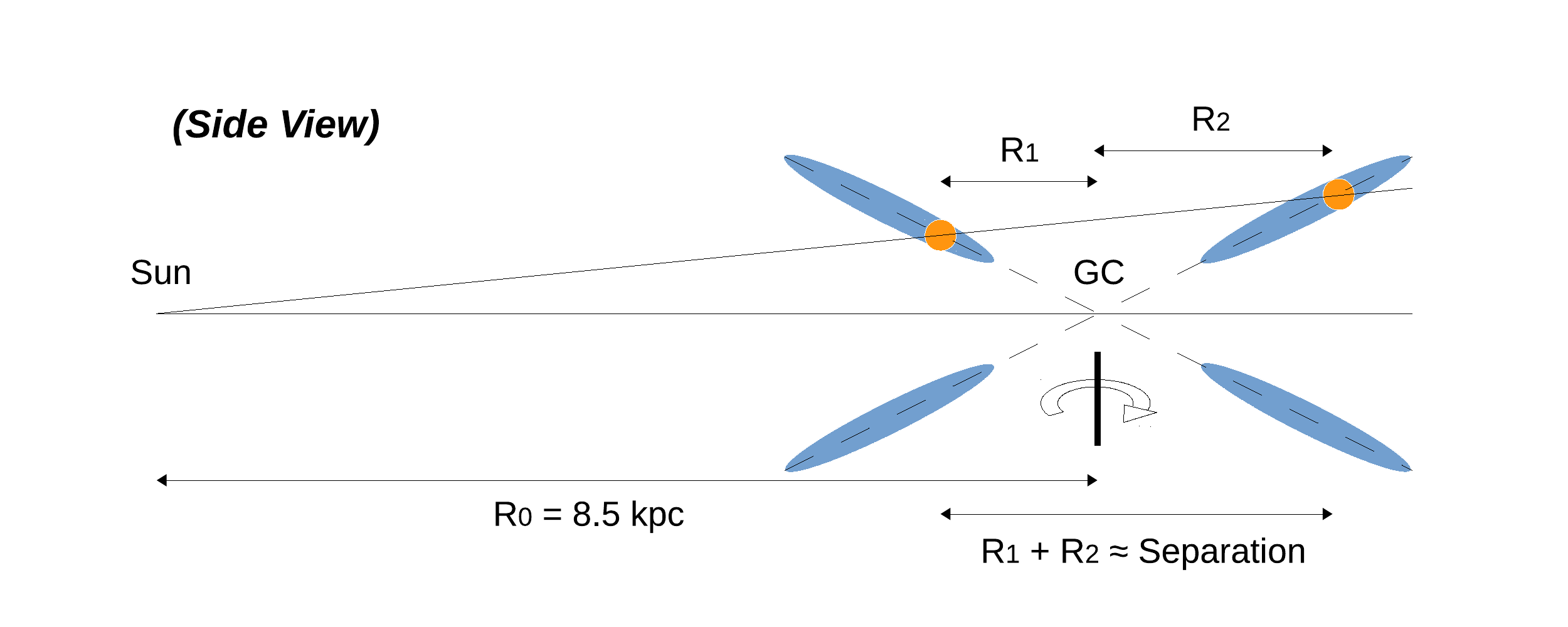}
\caption
{
Measuring the angular speed $\Omega$ from $\Delta\overline\mu_l^{\star}$ and the peak separation. The gray line represents a line of sight towards the bulge region, and the yellow dots show the positions of the two density peaks due to the X-shape (blue shaded regions). If X-shape is symmetric about the GC, we can obtain an angular speed $\Omega$ from the mean latitudinal proper motions near two peaks, and the separation between two peaks.
}
\label{fig:xshape}
\end{figure*}

The longitudinal proper motions at $l\sim0\degree$ reflect the azimuthal velocities at certain radii, and thus may be used to infer the bulge rotation. P13 derived ``the angular velocity of the Galactic bar'', i.e. the bar pattern speed $\Omega_p$, with the proper motion measured at the two arms of the X-shape, as illustrated in Figure \ref{fig:xshape}. We re-examine the P13 method with our model in the following. Assuming the Galactocentric radii of the near and far peaks are $R_1$ and $R_2$ respectively, and the mean azimuthal motion in the bar/bulge region corresponds to an angular velocity of $\Omega$, the longitudinal proper motions at the near and far peaks are
\begin{equation}
\mu_{1,l}^{\star} = \frac{R_1 \Omega - V_0}{R_0 - R_1}, \quad \mu_{2,l}^{\star} = \frac{-R_2 \Omega - V_0}{R_0 + R_2}.
\end{equation}
In practice, it is easier to measure the proper motion difference inside the same field. Thus, the mean difference of $\mu_l^{\star}$ between the two peaks can be written
\begin{equation}
\Delta\mu_{l}^{\star} = \mu_{1, l}^{\star} - \mu_{2, l}^{\star} = \frac{R_0(R_1 + R_2)\Omega - (R_1 + R_2)V_0}{(R_0 - R_1)(R_0 + R_2)}.
\end{equation}
Since $R_1$ and $R_2$ are small compared to $R_0$, let $\Delta R = R_1 + R_2$, given $\mu_{l, GC}^{\star} = -V_0 / R_0$, we find
\begin{equation}
\Omega \sim \frac{R_0}{\Delta R} \Delta \mu_{l}^{\star} - \mu_{l, GC}^{\star}
\end{equation}
and $\Delta R$ can be approximated with the separation between the two peaks along the line of sight.
This is essentially same as the equations in P13,
where they use the $I$-band magnitude difference of two red clumps ($\Delta I_{RC}$) to represent the distance between the two peaks ($\Delta R$).

With the procedure above, P13 obtained a bar pattern speed of $\Omega=-87.9\kmskpc$; this value is significantly higher than the pattern speed of this model ($38.5\kmskpc$, Molloy et al. 2015), and all current estimates of the pattern speed of the Galactic bar, which ranges from $30$ to $60\kmskpc$ (Gardner \& Flynn 2010; Gerhard 2011; Minchev et al. 2011; Wang et al. 2012, 2013; Long et al. 2013; Antoja et al. 2013). We derive an $\Omega$ of $\sim96\pm25\kmskpc$ at $(0\degree, 5\degree)$ for our model, a value similar to P13. If $\Omega$ indeed represents the bar pattern speed, it should be nearly constant across various $|b|$. 

Here we further test the P13 method by extending Equation 3 to different latitudes.
Table 2 shows the angular speed $\Omega$ derived from the peak separation and $\Delta\overline{\mu}_l^{\star}$ in our model.
Here ``peak particle'' means one quarter of particles closest to the identified peak at each side.
The separation between two peaks gradually increases towards higher latitudes, while for both peak particles and the overall samples, the variation of $\Delta\overline{\mu}_l^{\star}$ is insignificant. Consequently, the derived angular speed $\Omega$ decreases rapidly with $|b|$ in both samples (by nearly a factor of two from $|b| = 4\degree$ to $7\degree$).
As we have demonstrated, $\Omega$ varies significantly with $|b|$, so it cannot represent the bar pattern speed.

We suggest that $\Omega$ (Equation 3) reflects the mean azimuthal rotation velocity of the bar/bulge at certain radii and heights, which varies with the latitude and peak separation. At higher latitudes, the measured $\Omega$ probes the azimuthal rotation velocity at larger radii because the separation between the two peaks increases with $|b|$.

\begin{table*}
\label{tab:cylrot}
\begin{center}
\centering
\caption
{
Angular velocity of the Galactic bar/bulge from $\Delta\overline{\mu}_l^{\star}$
}
\begin{tabular}{l l l l l l}
\hline
Lat.    & Separation    & $\Delta\overline{\mu}_l^{\star}$  & $\Omega$                  & $\Delta\overline{\mu}_l^{\star}$ (Peak)   & $\Omega$ (Peak)       \\
($\deg$)& ($\kpc$)      & ($\masyr$)                        & ($\kmskpc$)               & ($\masyr$)                                & ($\kmskpc$)           \\
\hline
4.00    & 1.07          & 2.41                              & $121.56 \pm 37.31$        & 2.37                                      & $120.14 \pm 37.55$    \\
4.50    & 1.28          & 2.44                              & $107.12 \pm 29.96$        & 2.38                                      & $105.36 \pm 30.05$    \\
5.00    & 1.53          & 2.49                              & $96.38 \pm 24.55$         & 2.48                                      & $96.07 \pm 24.75$     \\
5.50    & 1.63          & 2.39                              & $89.62 \pm 22.71$         & 2.52                                      & $92.92 \pm 22.90$     \\
6.00    & 1.81          & 2.43                              & $84.65 \pm 20.13$         & 2.58                                      & $87.89 \pm 20.36$     \\
6.50    & 2.02          & 2.33                              & $76.96 \pm 17.81$         & 2.49                                      & $80.12 \pm 18.04$     \\
7.00    & 2.16          & 2.33                              & $74.09 \pm 16.46$         & 2.40                                      & $75.36 \pm 16.66$     \\
\hline
\end{tabular}
\end{center}
\textbf{Notes}:
\textit{Separation} indicates the distance between two peaks identified with kernel density estimator, $\Delta\overline{\mu}_l^{\star}$ stands for the difference of mean proper motion, and $\Omega$ is the derived angular velocity. \textit{Peak} is the $25\%$ particles nearest to the identified peaks. We assume a $5\%$ error in distance, and $10\%$ in $\mu$. The proper motion of the GC is $-5.46\masyr$ (assuming $V_0=220\kms$ at $R_0=8.5\kpc$).
\end{table*}

\begin{figure}[h!]
\centering
\includegraphics[width=0.75\columnwidth]{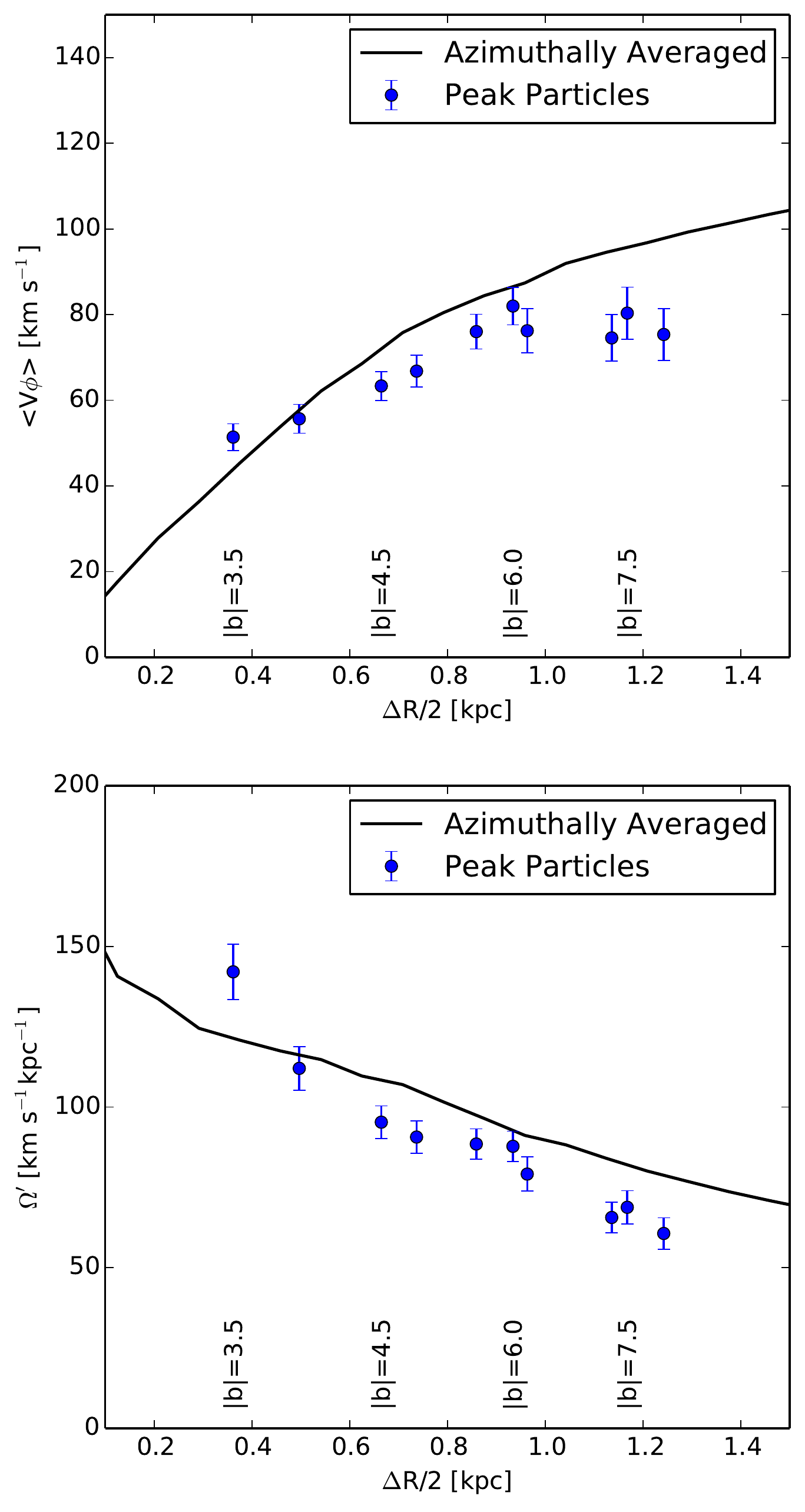}
\caption
{
Mean azimuthal rotation ($<V_{\phi}>$, top panel) and the corresponding angular velocity ($\Omega^\prime = <V_{\phi}> / R$, bottom panel) of the X-shaped peaks in the vertical $l=0\degree$ plane as a function of Gactocentric radius ($\sim \Delta R / 2$). In each panel, the blue points represent the peak particles, while the black solid line shows the azimuthally averaged values, regardless of the particle height $Z$. For the blue points, we also labelled the corresponding latitude $|b|$ where the particles are selected. Since $<V_{\phi}>$ decreases with increasing the height, at a given radius, the azimuthally (and vertically) averaged $<V_{\phi}>$ is smaller than the rotation velocity at lower $|Z|$ but larger than that at higher $|Z|$. Due to the X-shape geometry, the Galactocentric radius and the height of the peak particles (the blue dots) increase with $|b|$, resulting in larger $<V_{\phi}>$ at smaller radius and smaller $<V_{\phi}>$ at larger distance compared to the average value (the black line).
}
\label{fig:angular}
\end{figure}

\begin{figure*}[hb!]
\centering
\includegraphics[width=0.8\textwidth]{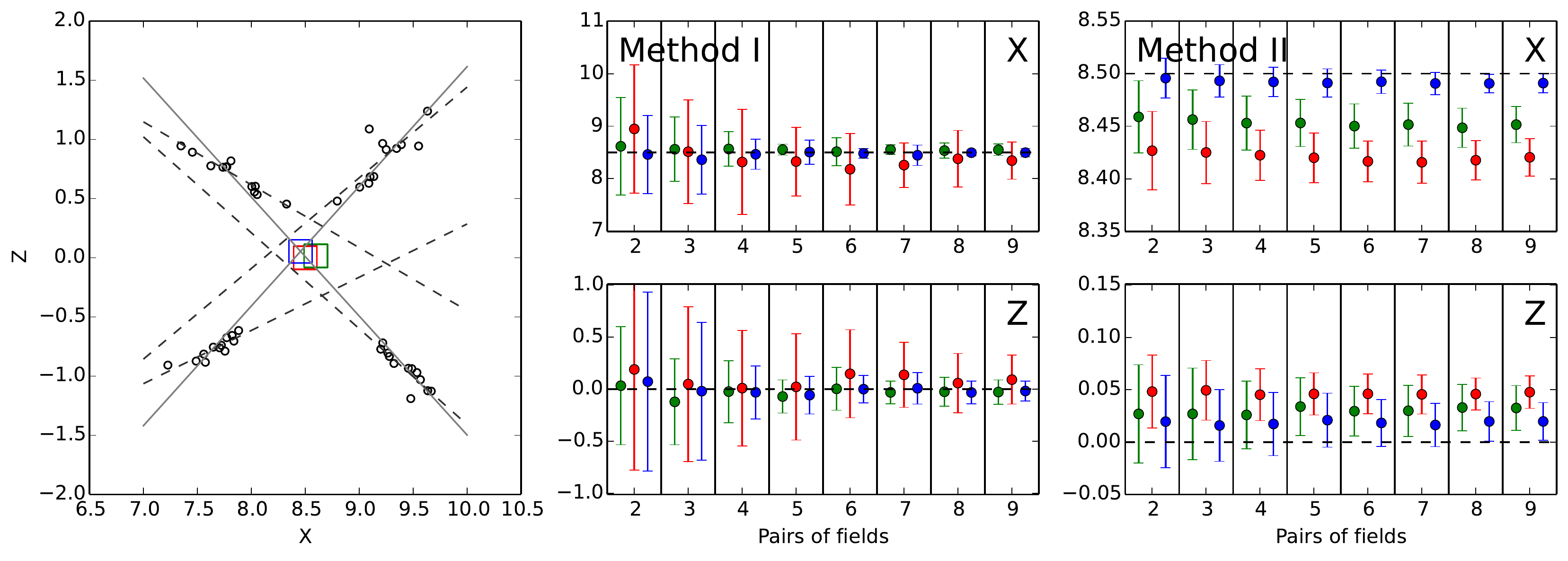}
\caption
{
The left panel shows an example of locating the Galactic center with the X-shape ($N=12$). Circles show the identified density peaks in the $l=0\degree$ plane, and the red box shows the real position of the GC. One can either fit the four arms and locate the GC using the intersection points (dashed lines and green box, Method I), or directly fit the diagonal pattern with two crossed lines to find the GC (solid lines and blue box, Method II). The rest two panels show the results from these two methods. The color dots indicate the fitted positions in two directions for each $N$, and the error bars show the standard deviation of the offset. The real position is marked with horizontal dashed lines. Their colors stand for the latitude range of the fields. For each $N$, from left to right, green, red and blue dots represent the entire, higher and lower latitudes, respectively.
}
\label{fig:gcfit}
\end{figure*}

To better understand this behavior, we compare the mean azimuthal velocity of peak particles with the azimuthally averaged rotational velocity of all the particles at the same Galactocentric radius, regardless of their height $Z$. We study a series of fields within $3.5\degree<|b|<8\degree$ (with an interval of $0.5\degree$) along $l=0\degree$. For each field, the average rotation velocity ($<V_{\phi}>$) and the corresponding angular velocity ($\Omega^\prime=<V_{\phi}> / R$)\footnote{Note that $<V_{\phi}>/R$ is different from the conventional angular frequency derived from the circular-speed curve.} are computed based on peak particles. An average radius ($\Delta R/2$) for each pair is obtained by averaging the Galactocentric radius of the two peaks. Of course, this radius increases for higher Galactic latitudes. Therefore, $\Omega$ from Equation 3 at different Galactic latitudes can be naturally translated into a function of radius. Figure 14 shows $<V_{\phi}>$ and $\Omega^\prime$ as a function of the Galactocentric radius  $\Delta R/2$. The solid line represents the azimuthally averaged $<V_{\phi}>$. Apparently, $<V_{\phi}>$ and $\Omega^\prime$ of peak particles (blue dots) agree very well with the azimuthally averaged values of all the particles (black solid line). Therefore, $\Omega$ in Equation 3 actually measures the azimuthal streaming motions ($<V_{\phi}>$), rather than the bar pattern speed.

\subsection{Locating the Galactic Center with the X-shape}

Since the X-shape is symmetric in the $l\sim0\degree$ plane, it can be used to locate the position of the GC (McWilliam \& Zoccali 2010; G14). The GC can be located by fitting the peak positions in the vertical $l=0\degree$ plane with the X-shape geometry. G14 compared two ways to locate the GC position, namely fitting the four arms of the X-shape, versus fitting the X-shape with two crossed lines. They found that Method II yields an accuracy better than $5\%$ in most cases, even if the peak distance has an uncertainty of $20\%$. However, the sample limitations (e.g. latitude coverage, number of fields, etc.) may affect the results. We try to locate the GC in our model with these two methods, and further beyond G14, we test their accuracies under different latitude coverages and total number of fields.

We select $N$ pairs of fields along the minor axis of the bulge ($l=0\degree$), with $N$ fields above and $N$ fields below the Galactic plane respectively. Fields are randomly selected in the latitude range where the X-shape can be clearly identified in this model ($3\degree<|b|<8\degree$). In each field, the two peaks are identified with the same procedure described in Section 2.1. We then fit the arms of the X-shape with four straight lines, and find their intersection points (Method I). In the ideal case, two of their four intersection points are right above/below the GC (i.e. on the vertical axis of the Galaxy), and the other two points are exactly on the Galactic plane. Thus the GC position can be located from the Galactic vertical axis, and the Galactic plane. Meanwhile, we try to locate the GC by directly fitting the X-shape with two crossed lines (Method II). The two methods are illustrated in the left panel of Figure \ref{fig:gcfit}. We run a series of tests with different $N$ to demonstrate the effect of limited field number. To illustrate the impact of limited latitude coverage, in another series of tests, we also change the latitude range to $3\degree<b<5.5\degree$ (``lower'') or $5.5\degree<b<8\degree$ (``higher''), and compare the results with the case of full latitude coverage ($3\degree<b<8\degree$).

We find that, fitting the X-shape with two crossed lines (Method II, right panel of Figure \ref{fig:gcfit}) yields better results compared to fitting the four individual arms (Method I, middle of Figure \ref{fig:gcfit}), which agrees with the conclusion in G14. Meanwhile, increasing the number of fields leads to smaller uncertainties for both methods. We also find that, the accuracy of fitting depends on the latitude range of the fields used. The position of the GC can be better constrained with the fields at lower latitudes, while the fields at higher latitudes give the worst results. In principle, fields at higher latitudes put stronger constraints on the GC position. However, since the X-shape is well-defined at lower latitudes ($3\degree<l<5.5\degree$) due to better number statistics (also seen in the upper right panel of Figure 1), the fitting results from these regions are more accurate compared to higher latitudes ($5.5\degree<l<8\degree$).

Fitting the X-shaped overdensity is a novel and accurate way to determine the position of the Galactic center. However, it still has some drawbacks. First it relies on the identification of peaks in the apparent magnitude distributions of the sample, which could have large uncertainties in real observations due to the presence of scatters in the intrinsic luminosity, source incompleteness and sample contamination, etc. In the conversion from magnitude to distance, reddening and extinction could also contribute to the uncertainties.

Even if the uncertainties due to photometric distance determination can be greatly reduced with geometric distance measurements like \textit{Gaia} does, there remains uncertainties in Method II. Ideally, if the peaks at the near side have similar heights as the far side, fitting two diagonal lines should trace the exact position of GC. However, due to the projection effect, the peaks identified at the near side are closer to the disk plane than the far side. Consequently, the intersection of the two best-fit crossed lines tends to be closer to the Sun, leading to a slight systematic underestimation of the GC distance. Since the projection effect is stronger at higher latitude, fields at lower latitudes are expected to yield better results than those at higher latitudes, which agrees with the right panel of Figure \ref{fig:gcfit}. In reality, the number of observational fields are usually not symmetric above and below the Galactic plane, which may also affect the best-fit crossed lines to locate the GC.

\section{CONCLUSIONS}

We explore the kinematic properties of the X-shaped Milky Way bulge with an $N$-body model that well represents the observed morphology and dynamics of the Galactic bulge. From the solar perspective, the two density peaks on the sightlines due to the X-shape are identified using a Gaussian kernel density estimator with adaptive kernel size. In this model, the distance distributions appear to be double-peaked above $|b|\sim3\degree$ along $l=0\degree$, and the increasing separation between two peaks with $|b|$ agrees with previous observations. As a result of a tilted bar, the height ratio of the two peaks (far to near), as well as their mean distance to the Sun, features a longitudinal gradient at higher latitudes.

We study the line-of-sight velocity and proper motion distributions in fields ($b=2\degree,4\degree,6\degree$ and $8\degree$) along $l=0\degree$, with the opposite fields properly scaled and added to increase number statistics. To illustrate the possible imprints of the bar/X-shape, the particles are divided into two groups (near and far) according to their distances to the Sun compared to $R_0$. In these fields, the line-of-sight velocity distributions significantly differ between the near and far sides; the near side shows an excess of approaching particles, while a corresponding excess of receding particles exists in the far side. Due to the rotation of the bar/bulge, the longitudinal proper motions at the near and far sides peak at the positive and negative sides of $\mu_{l,GC}^{\star}$, respectively. The measured mean differences of $\mu_l^{\star}$ between the two sides are comparable to the proper motion dispersion there, indicating the fast cylindrical rotation of the bar/bulge. The latitudinal proper motions at the two sides barely differ in their mean values. We note that the differences of $V_{\mathrm{los}}$ and $\mu_l^{\star}$ between the two sides can be qualitatively explained with the coherent alignment of bar-supporting orbits.

We test if the larger proper motion dispersion measured at the near side (compared to the far side) is due to the closer distance. In our model, only the dispersion ratio of $\mu_l^{\star}$ between the two sides (near to far) is marginally consistent with the distance ratio of the two peaks (far to near). The results here indicate that the systematically larger dispersion at the near side is not only due to the closer distance, but also contributed by the differences in their intrinsic velocity distributions.

We also map the mean line-of-sight velocity, mean proper motion and their corresponding dispersions across the bulge region, both with and without distinguishing the near and far sides of the Galactic bar/bulge. The mean $V_{\mathrm{los}}$ at the two sides features strong cylindrical rotation, and the position of the zero-velocity line with respect to $l=0\degree$ agrees with the coherent alignment or bar-supporting orbits. The mean difference of $V_{\mathrm{los}}$ between the near and far sides shows a shallow minimum along $l=0\degree$, which is suggested as the signature of an X-shaped bulge (G14). Without distinguishing the near and far sides, $\overline{V}_{\mathrm{los}}$ is symmetric with respect to $l=0\degree$, which barely shows the effect of the tilted bar as reported in Zoccali et al. (2014). For the mean $\mu_l^{\star}$ at the two sides, due to the coherent alignment of bar-supporting orbits, the extrema shift to the two opposite sides of $l=0\degree$ respectively. At lower latitudes, the displacements of the extrema are larger than that at higher latitudes. This may be due to the buckled inner region of the bar dominating at higher latitudes, while the long and thin component of the bar dominates at lower latitudes. Inside $-1\degree<l<1\degree$, $b = -5\degree$, the mean differences of $\mu_l^{\star}$ and $\mu_b$ between the two sides are roughly symmetric with respect to $l=0\degree$. The breaks of $\Delta\overline{\mu}_l^{\star}$ and $\Delta\overline{\mu}_b$ reported in P13 are absent in our model.

For their dispersions, $\sigma_{\mathrm{los}}$, $\sigma_l$ and $\sigma_b$ at both near and far sides reveal the kinematically hot central region. The dispersion ratio is saddle-shaped at both sides, with a diagonal pattern of the cross-correlation factor, indicating that the proper motion anisotropy is tilted towards the GC. Without distinguishing the near and far sides of the bar/X-shape, the dispersion of $V_{\mathrm{los}}$ shows a vertically elongated peak similar to Zoccali et al. (2014). Meanwhile, the proper motion dispersion is larger in the longitudinal direction, and the dispersion ratio ($\sigma_l/\sigma_b$) increases away from $l=0\degree$. We also observe a diagonal pattern of the cross-correlation factor. Compared to the proper motion measurements in Rattenbury et al. (2007), this model shows slightly smaller dispersion but stronger anisotropy.

We also study the orbital motion inside the X-shape using the correlation of velocity components. In $(0\degree,6\degree)$, for $\mu_b$ vs. $V_{\mathrm{los}}$ and $\mu_b$ vs. $\mu_l^{\star}$, both the significance of the correlations (Pearson's r value) and the slope from linear regressions reveal slightly stronger velocity anisotropy near the X-shape. Except for $\mu_b$ vs. $\mu_l^{\star}$ at the near side, the correlation is weakest for the peak sample. We note that at both sides, the direction of the anisotropy roughly coincides with the case of downward-tilt banana orbits (for $b>0\degree$). However, the underlying orbital family can not be clearly distinguished here.

With this model, we extend the azimuthal rotation velocity measured in P13 with the X-shape kinematics to various Galactic latitudes. We find that the angular velocity decreases at higher Galactic latitudes, inconsistent with being the bar pattern speed. The angular velocity derived from the proper motion difference between the two sides of the X-shape actually measures the azimuthal streaming motions, similar to the mean rotation profile $<V_{\phi}>(R)$. 

The X-shape is thought to be able to constrain the position of the Galactic center due to its symmetry. We test two methods to do this proposed in Gardner et al. (2014) with our model, namely fitting the four arms of the X-shape to find the center and diagonally joining the four arms by fitting the X-shape with two crossed lines. We demonstrate that a global-fitting yields better results than fitting the four arms individually. Also, the fields at lower latitudes provide stronger constraints on the GC position due to better number statistics there.

\section*{ACKNOWLEDGEMENTS}

We thank the anonymous referee for constructive comments that help to improve the paper, especially Section 3.5. The research presented here is partially supported by the 973 Program of China under grant no. 2014CB845700, by the National Natural Science Foundation of China under grant nos.11333003, 11322326, 11403072, and by the Strategic Priority Research Program  ``The Emergence of Cosmological Structures'' (no. XDB09000000) of the Chinese Academy of Sciences. ZYL is grateful for the support from Shanghai Sailing Program (No. 14YF1407700). This work made use of the facilities of the Center for High Performance Computing at Shanghai Astronomical Observatory. Hospitality at APCTP during the 7th Korean Astrophysics Workshop is kindly acknowledged.

\begin{appendix}

\section*{Appendix A: Calibrated Silverman's Test}

In this section, we present a brief description of the Silverman's test used in Section 3.1. The Silverman's test (Silverman 1981) is based on the kernel density estimation of the sample. For a univariate sample $X_1$, $\ldots$, $X_n$ drawn from an unknown probability distribution function (PDF) $f$, its kernel density estimator (KDE) is
$$ \hat{f}(x;h) = \frac{1}{nh} \sum_{i=1}^{n} K(\frac{x - X_i}{h}),$$
where $x$ is the random variable, $K$ is the kernel function which is assumed to be a Gaussian distribution function, and $h$ is the kernel size. Here $\hat{f}$ serves as an approximation of the unknown original PDF $f$. It resembles a continuous version of the histogram, and $h$ corresponds to the ``bin size''. The kernel size $h$ controls the amount of smoothing; larger $h$ leads to fewer peaks in the KDE with the Gaussian kernel.

To test the null hypothesis that $f$ has $k$ or fewer modes ($H_0$) against the alternative one that $f$ has more than $k$ modes ($H_1$), we use the critical kernel size of $k$ modes
$$ h_k = \inf \{ h; \hat{f}(\cdot ;h) \mathrm{\;has\; at\; most\;} k \mathrm{\;modes} \}.$$
Large $h_k$ rejects the null hypothesis, indicating that there are more than $k$ peaks. In other words, the $k$-critical kernel size is the smallest kernel size that makes a KDE of $k$ peaks, for a particular sample. If the sample has more than $k$ modes, a large $h_k$ is required to over-smooth the KDE, and reduce the peak number to $k$. On the contrary, if the sample itself has less than $k$ modes, only an under-smoothed KDE can result in more than $k$ peaks, which requires a small $k$-critical kernel size.

For a particular $k$-mode hypothesis, now we address the statistical significance of the $k$-critical kernel size from the data ($h_{k,0}$). In principle, one can draw a series of samples with equal size $n$ from $f$, find their $k$-critical kernel sizes ($h_k^{\star}$), and compare the critical kernel size from the data to those from resampling the true PDF $f$ using the $\alpha$ level test:
$$ \mathrm{pr}\{h_k^{\star} \leq h_{k,0} | X_1, \ldots, X_n \mathrm{\;is\;drawn\;from\;} f\} \geq 1 - \alpha, $$
where $\mathrm{pr}$ represents the conditional probability, and $\alpha$ is the level of significance. Since the peak number decreases with increasing the kernel size, we can address the significance level of $h_{k,0}$ by accounting for the peak number of $\hat{f}(x;h_{k,0})$ from resampling:
$$ \mathrm{pr}\{h_k^{\star} > h_{k,0}\} = \mathrm{pr}\{\hat{f}(\cdot;h_{k,0}) \mathrm{\;has\; more\; than\;} k \mathrm{\;modes\;} |$$
$$ \{X_1, \ldots, X_n\} \mathrm{\;is\; drawn\; from\;} f \}.$$

However, the true density $f$ is always unknown. In order to evaluate the significance of $h_{k,0}$, Silverman (1981) generated the samples using the method in Efron (1979): 
$$ Y_i = (X_{j} + h_{k,0} \epsilon_i ) / \sqrt{1 + h_{k,0}^2 / \sigma^2},$$
where $X_i$ are sampled uniformly from the original data allowing replacement, $\sigma^2$ is the variance of the sample, $h_{k,0}$ is the $k$-critical kernel size from the data, and $\epsilon_i$ is a series of random numbers from a standard normal distribution. Here $Y_i$ actually samples a representative PDF $f_0$, which is at most $k$-modal, and has the same variance as the sample. Therefore, the significance level of $h_{k,0}$ can be conservatively estimated with such a bootstraping procedure.

Besides its conservative nature, the Silverman's test is also not asymptotically accurate. One way to improve the accuracy of Silverman's multimodality test is presented in Hall \& York (2011), by using a bootstraping distribution function
$$ \hat{G}_n(\lambda) = \mathrm{pr}\{ \hat{h}^{\star}_k / h_{k,0} \leq \lambda | \{ X_i, \ldots, X_n \} \}.  $$
As suggested in Hall \& York (2011), there exists a unique absolute constant $\lambda_{\alpha}$ such that the bootstrap test is asymptotically correct:
$$ \mathrm{pr}\{\mathrm{pr}\{\hat{h}^{\star}_k / h_{k,0} \leq \lambda_{\alpha} | \{ X_i, \ldots, X_n \} \} \geq 1 - \alpha\}$$
$$\to \mathrm{pr}\{\hat{G}(\lambda_{\alpha}) \geq 1 - \alpha\} = \alpha, $$
and the constant $\lambda_{\alpha}$ can be calculated either analytically, or using Monte Carlo simulations. In addressing the significance level of the double-peaked feature ($p_1 = 1 - \alpha$), we use the tabular value given in Hall \& York (2011).

\end{appendix}


\begin{thebibliography}

\bibitem[Athanassoula(2005)]{2005CeMDA..91....9A} Athanassoula, E. \ 2005, Celestial Mechanics and Dynamical Astronomy, 91, 9

\bibitem[Binney et al.(1997)]{1997MNRAS...288..365B} Binney, J., Gerhard, O., Spergel, D., \ 1997, \mnras, 288, 365

\bibitem[Bissantz \& Gerhard(2002)]{2002MNRAS.330..591B} Bissantz, N., Gerhard, O., \ 2002, \mnras, 330, 591

\bibitem[Blitz \& Spergel(1991)]{1991ApJ...379..631B} Blitz, L., Spergel, D.~N., \ 1991, \apj, 379, 631

\bibitem[Bureau et al.(2006)]{2006MNRAS.370..753B} Bureau, M., Aronica, G., Athanassoula, E., et al. \ 2006, \mnras, 370, 753

\bibitem[Cao et al.(2013)]{2013MNRAS.434..595C} Cao, L., Mao, S., Nataf, D., \ 2013, \mnras, 434, 595

\bibitem[Combes \& Sanders(1981)]{1981A&A....96..164C} Combes, F., Sanders, R.~H., \ 1981, \aap, 96, 164

\bibitem[De Propris et al.(2011)]{2011ApJ...732L..36D} De Propris, R., Rich, R.~M., Kunder, A., et al., \ 2011, \apjl, 732, 36

\bibitem[de Vaucouleurs(1964)]{1964IAUS...20..195D} de Vaucouleurs, G. \ 1964, Proceedings of the International Astronomical Union Symposium No. 20, 195

\bibitem[Dwek et al.(1995)]{1995ApJ...445..716D} Dwek, E., Arendt, R.~G., Hauser, M.~G., et al.\ 1995, \apj, 445, 716

\bibitem[Efron(1979)]{} Efron, B., \ 1979, Ann. Statist., 7, 1

\bibitem[Englmaier \& Gerhard(1999)]{1999MNRAS.304..512E} Englmaier, P., Gerhard, O., \ 1999, \mnras, 304, 512

\bibitem[Fux(1999)]{1999A&A...345..787F} Fux, R., \ 1999, \aap, 345, 787

\bibitem[Gardner et al.(2014)]{2014MNRAS.438.3275G} Gardner, E., Debattista, V.~P., Robin, A.~C., et al. \ 2014, \mnras, 438, 3275

\bibitem[Hall \& York(2001)]{} Hall, P., York, M., \ 2001, Statistica Sinica, 11, 515

\bibitem[Howard et al.(2008)]{2008ApJ...688.1060H} Howard, C.~D., Rich, R.~M., Reitzel, D.~B., et al.\ 2008, \apj, 688, 1060

\bibitem[Howard et al.(2009)]{2009ApJ...702L.153H} Howard, C.~D., Rich, R.~M., Clarkson, W., et al.\ 2009, \apjl, 702, L153

\bibitem[Kunder et al.(2012)]{2012AJ....143...57K} Kunder, A., Koch, A., Rich, R.~M., et al.\ 2012, \aj, 143, 57

\bibitem[Koz\l owski et al.(2006)]{2006MNRAS.370..435K} Koz\l owski, S., Wo\'zniak, P.~R., Mao, S., et al. \ 2006, \mnras, 370, 435

\bibitem[Li \& Shen (2012)]{2012ApJ...757L...7L} Li, Z.-Y. \& Shen, J., \ 2012, \apjl, 757, L7

\bibitem[Li et al.(2014)]{2014ApJ...785L..17L} Li, Z.-Y., Shen, J., Rich, R.~M., Kunder, A. \& Mao, S., \ 2014, \apjl, 785, 17 

\bibitem[L\"{u}tticke et al.(2000)]{2000A&AS..145..405L} L\"{u}tticke, R., Dettmar, R.-J. \& Pohlen, M., \ 2000, \aaps, 145, 405

\bibitem[McWilliam & Zoccali(2010)]{2010ApJ...724.1491M} McWilliam, A., Zoccali, M., \ 2010, \apj, 724, 1491

\bibitem[Molloy et al.(2015)]{2014arXiv1412.4689M} Molloy, M., Smith, M.~C., Shen, J., Evans, N.~W., \ 2015, \apj, in press, (arXiv:1412.4689)

\bibitem[Nataf et al.(2010)]{2010ApJ...721L..28N} Nataf, D.~M., Udalski, A., Gould, A, et al., \ 2010, \apjl, 721, 28

\bibitem[Nataf et al.(2015)]{2015MNRAS.447.1535N} Nataf, D.~M., Udalski, A., Skowron, J., et al., \ 2015, MNRAS, 447, 1535

\bibitem[Ness et al.(2012)]{2012ApJ...756...22N} Ness, M., Freeman, K., Athanassoula, E., et al.\ 2012, \apj, 756, 22

\bibitem[Nidever et al.(2012)]{2012ApJ...755L..25N} Nidever, D.~L., Zasowski, G., Majewski, S.~R., et al.\ 2012, \apjl, 755, L25

\bibitem[Patsis et al.(2002)]{2002MNRAS.337..578P} Patsis, P.~A., Skokos, Ch. \& Athanassoula, E, \ 2002, \mnras, 337, 578

\bibitem[Patsis \& Katsanikas(2014a)]{2014MNRAS.445.3525P} Patsis, P.~A., Katsanikas, M. \ 2014, \mnras, 445, 3525

\bibitem[Patsis \& Katsanikas(2014b)]{2014MNRAS.445.3546P} Patsis, P.~A., Katsanikas, M. \ 2014, \mnras, 445, 3546

\bibitem[Pfenniger \& Friedli (1991)]{1991A&A...252...75P} Pfenniger, D., Friedli, D. \ 1991, \aap, 252, 75

\bibitem[Poleski et al.(2013)]{2013ApJ...776...76P} Poleski, R., Udalski, A., Gould, A., et al. \ 2013, \apj, 776, 76

\bibitem[Portail et al.(2015a)]{2015MNRAS.448..713P} Portail, M., Wegg, C., Gerhard, O. \& Martinez-Valpuesta, I, \ 2015a, \mnras, 448, 713

\bibitem[Portail et al.(2015b)]{2015arXiv150307203P} Portail, M., Wegg, C., \& Gerhard, O. \ 2015b, \mnras, in press, (arXiv:1503:07203)

\bibitem[Raha et al.(1991)]{1991Natur.352..411R} Raha, N., Sellwood, J.~A., James, R.~A., \& Kahn, F.~D. \ 1991, \nat, 352, 411

\bibitem[Rattenbury et al.(2007)]{2007MNRAS.378.1064R} Rattenbury, N.~J., Mao, S., Sumi, T., Smith, M.~C. \ 2007,  \mnras, 378, 1064

\bibitem[Saito et al.(2011)]{2011AJ....142...76S} Saito, R.~K., Zoccali, M., McWilliam, A., et al. \ 2011, \aj, 142, 76

\bibitem[Shen et al.(2010)]{2010ApJ...720L..72S} Shen, J., Rich, R.~M., Kormendy, J., et al.\ 2010, \apjl, 720, L72

\bibitem[Shen(2014)]{2014IAUS..298..201S} Shen, J., \ 2014, in Proceedings of IAU Symposium 298, 201

\bibitem[Silverman(1981)]{} Silverman, B.~W., \ 1981, J. R. Statist. Soc. B, 43, 97

\bibitem[Soto et al.(2012)]{2012A&A...540A..48S} Soto, M., Kuijken, K., Rich, R.~M. \ 2012, \aap, 540, A48

\bibitem[Stanek et al.(1994)]{1994ApJ...429L..73S} Stanek, K.~Z., Mateo, M., Udalski, A., et al. \ 1994, \apjl, 429, 73

\bibitem[V\'asquez et al.(2013)]{2013A&A...555..A91} V\'asquez, S., Zoccali, M., Hill, et al.\ 2013, \aap, 555, A91

\bibitem[Whitemore \& Bell (1988)]{1988ApJ...324..741W} Whitmore, Bradley~C., Bell, M. \ 1988, \apj, 324, 741

\bibitem[Williams et al.(2011)]{2011MNRAS.414.2163W} Williams, M.~J., Zamojski, M.~A., Bureau, M., et al. \ 2011, \mnras, 414, 2163

\bibitem[Wegg \& Gerhard (2013)]{2013MNRAS.435.1874W} Wegg, C., Gerhard, O., \ 2013, \mnras, 435, 1874

\bibitem[Zoccali et al.(2014)]{2014A&A...562A..66Z} Zoccali, M., Gonzalez, O.~A., Vasquez, S., et al. \ 2014, \aap, 562, 66

\end{thebibliography}
\end{document}